\documentclass[a4paper,fleqn,twocolumn,margin=1in]{aastex63}
\pdfoutput=1
\usepackage{float}
\usepackage{graphicx}
\usepackage{ae,aecompl} 
\usepackage{enumitem} 
\usepackage{amssymb} 
\usepackage{amsmath} 
\usepackage{hyperref} 
\usepackage[lofdepth,lotdepth,caption=false]{subfig}

\def\apjl{\rm ApJL} 
\def\apjs{\rm ApJS}

\def\aap{\rm AAP}

\def\gax{\mathrel{\raise.3ex\hbox{$>$}\mkern-14mu\lower0.6ex\hbox{$\sim$}}} 
\def\lax{\mathrel{\raise.3ex\hbox{$<$}\mkern-14mu\lower0.6ex\hbox{$\sim$}}} 
\def\gtorder{\mathrel{\raise.3ex\hbox{$>$}\mkern-14mu 
             \lower0.6ex\hbox{$\sim$}}} 
\def\ltorder{\mathrel{\raise.3ex\hbox{$<$}\mkern-14mu 
             \lower0.6ex\hbox{$\sim$}}}

\def\varpiseis{\varpi_{\mathrm{seis}}}

\def\varpigaia{\varpi_{Gaia}}
\def\hatvarpigaia{\hat{\varpi}_{Gaia}}

\def\muhz{\mu\mathrm{Hz}}
\def\numax{\nu_{\mathrm{max}}}
\def\hatnumax{\hat{\nu}_{\mathrm{max}}}
\def\dnu{\Delta \nu}
\def\hatdnu{\hat{\Delta \nu}}

\def\teff{T_{\mathrm{eff}}}
\def\hatteff{\hat{T}_{\mathrm{eff}}}

\def\teffsun{T_{\mathrm{eff,} \odot}}

\def\rsun{R_{\odot}}

\def\msun{M_{\odot}}
\def\numaxsun{\nu_{\mathrm{max,} \odot}}
\def\dnusun{\Delta\nu_{\odot}}

\def\invrscal{R^{-1}_{\mathrm{seis}}}
\def\rscal{R_{\mathrm{seis}}}

\def\hatvarpigaia{\hat{\varpi}_{Gaia}}
\def\hatvarpigaiai{\hat{\varpi}_{Gaia,i}}
\def\varpigaiaj{\varpi_{Gaia,j}}
\def\varpigaiai{\varpi_{Gaia,i}}

\def\rscal{R_{\mathrm{seis}}}

\def\varpiast{\varpi_{\mathrm{seis}}}
\def\muas{\mu \mathrm{as}}
\def\mas{\mathrm{mas}}
\def\nueff{\nu_{\mathrm{eff}}}
\def\hatnueff{\hat{\nu}_{\mathrm{eff}}}
\def\ks{K_{\mathrm{s}}}
\def\hatvarpiast{\hat{\varpi}_{\mathrm{seis}}}
\def\hatvarpiasti{\hat{\varpi}_{\mathrm{seis},i}}
\def\varpiasti{{\varpi}_{\mathrm{seis},i}}

\begin{document}
\title{Testing the radius scaling relation with {\it Gaia} DR2 in the
  {\it Kepler} field}
\author{Joel C. Zinn}
\affiliation{School of Physics, University of New South Wales, Barker
  Street, Sydney, NSW 2052, Australia}
\affiliation{Department of Astronomy, The Ohio State University, 140
  West 18th Avenue, Columbus OH 43210, USA}
\affiliation{Kavli Institute for Theoretical Physics, University of
  California, Santa Barbara, CA 93106, USA}
\author{Marc H. Pinsonneault}
\affiliation{Department of Astronomy, The Ohio State University, 140 West
  18th Avenue, Columbus OH 43210, USA}
\author{Daniel Huber}
\affiliation{ Institute for Astronomy, University of Hawai`i, 2680 Woodlawn Drive, Honolulu, HI 96822, USA}
\author{Dennis Stello}
\affiliation{School of Physics, University of New South Wales, Barker
  Street, Sydney, NSW 2052, Australia}
\affiliation{Sydney Institute for Astronomy (SIfA), School of Physics,
  University of Sydney, NSW 2006, Australia}
\affiliation{Stellar Astrophysics Centre, Department of Physics and
Astronomy, Aarhus University, Ny Munkegade 120, DK-8000
Aarhus C, Denmark}
\affiliation{Center of Excellence for Astrophysics in Three Dimensions
(ASTRO-3D), Australia}
\author{Keivan Stassun}
\affiliation{Vanderbilt University, Department of Physics \&
Astronomy, 6301 Stevenson Center Ln., Nashville, TN 37235, USA}
\affiliation{Fisk University, Department of Physics, 1000 17th Ave. N., Nashville,
TN 37208, USA}
\author{Aldo Serenelli}
\affiliation{Institute of Space Sciences (ICE, CSIC) Campus UAB, Carrer de Can Magrans, s/n, 08193, Bellaterra, Spain}
\affiliation{Institut d'Estudis Espacials de Catalunya (IEEC), C/Gran Capita, 2-4, 08034, Barcelona, Spain}

\correspondingauthor{Joel C. Zinn}
\email{j.zinn@unsw.edu.au}
\begin{abstract}
We compare radii based on {\it Gaia} parallaxes to asteroseismic                                            
scaling relation-based radii of $\sim 300$ dwarfs \& subgiants and                                         
$\sim 3600$ first-ascent giants from                                                                       
the {\it Kepler} mission. Systematics due to                                                                
temperature, bolometric correction, extinction, asteroseismic radius,                                       
and the spatially-correlated {\it Gaia} parallax zero-point, contribute to a $2\%$                          
systematic uncertainty on the {\it Gaia}-asteroseismic radius                                               
agreement. We find that dwarf and giant scaling                                                             
radii are on a parallactic scale at the $-2.1 \% \pm 0.5 \% {\rm \                                          
(rand.)} \pm 2.0\% {\rm \ (syst.)}$ level (dwarfs) and $+1.7\% \pm                                          
0.3\% {\rm \ (rand.)} \pm 2.0\% {\rm (syst.)}$ level (giants),                                              
supporting the accuracy and precision of scaling relations in this                                          
domain. In total, the $2\%$ agreement that we find holds for stars                                          
spanning radii between $0.8\rsun$                                                                           
and $30 \rsun$. We do, however, see evidence for \textit{relative}                                          
errors in scaling radii between dwarfs and giants at the $4\% \pm 0.6\%$                                            
level, and find evidence of departures from simple scaling                                                  
relations for radii above $30 \rsun$. Asteroseismic masses for very metal-poor stars                        
are still overestimated relative to astrophysical priors, but at a                                          
reduced level.  We see no trend with metallicity in radius agreement                                        
for stars with $-0.5 <$ [Fe/H] $< +0.5$. We quantify the spatially-correlated parallax errors in the {\it                          
Kepler} field, which globally agree with the {\it Gaia} team's published covariance model. We provide {\it Gaia} radii,                                                                    
corrected for extinction and the {\it Gaia} parallax zero-point for                                         
our full sample of $\sim 3900$ stars, including dwarfs, subgiants, and                                      
first-ascent giants.
\end{abstract}

\section{Introduction}
\label{sec:intro}

Stellar astrophysics is in the midst of a radical transformation.
Massive surveys using a variety of tools --- time domain, astrometric,
photometric, and spectroscopic --- are yielding a wealth of information
about stars.  This treasure trove is not merely far larger than prior
data sets; it also contains fundamentally new information. This is
particularly true for fields studied by the \textit{Kepler} satellite,
where we have detected stellar oscillations in hundreds of stars near
the main sequence turnoff (e.g., \citealt{chaplin+2011c}) and tens of
thousands of evolved giant stars (e.g., \citealt{yu+2018}). The focus
of this paper is to test the accuracy and precision of radii that have been derived from \textit{Kepler}
asteroseismology.

Virtually all cool stars excite solar-like oscillations. Most stellar
population studies distill the information in the oscillation spectrum
down to two characteristic frequencies: the frequency of maximum power, $\numax$,
and the large frequency spacing, $\dnu$.  These can be related to
stellar mass and radius through scaling relations. The frequency of
maximum power is related to the acoustic cut-off frequency, and by
extension the surface gravity and effective temperature
\citep{brown+1991,kjeldsen&bedding1995}. The large frequency spacing is related to the mean
density, which can be demonstrated with asymptotic pulsation theory \citep{tassoul1980,christensendalsgaard1993}.
In simple scaling relations one therefore solves for two equations in
two unknowns, yielding asteroseismic masses and radii as a function of
$\teff$ and the asteroseismic parameters.  With the addition of
abundances from high-resolution spectra, stellar ages can also be
derived.  The APOGEE-{\it Kepler}, or APOKASC, collaboration was set up to
take advantage of this exciting prospect.

APOGEE uses an infrared spectrograph with R =  22,500 used in
combination with the SDSS 2.5-m telescope \citep{gunn+2006}.
The APOGEE \citep{majewski+2010} temperature scale has been calibrated
to agree with the IRFM temperature scale \citep{holtzman+2015}, and
the temperatures have recently been re-calibrated to correct for
evolutionary state- and metallicity-dependent trends in the most
recent data release, DR14 \citep{holtzman+2018}.

\cite{pinsonneault+2014} combined APOGEE spectroscopic temperatures
and metallicities with asteroseismic information for nearly 2000
giants in a forward-modeling exercise that reported typical precisions
in mass and radius of $12\%$ and $5\%$. This work represented the
largest application of asteroseismology to determine fundamental
stellar quantities, and clearly demonstrated the use of
asteroseismology in stellar populations work: the mass, radius, and
surface gravity of thousands of stars could be shown to be reasonable
and nominally extremely precise. Nevertheless, there was room for
improvements. For instance, it seemed evident that there were evolutionary
state--dependent systematics that could not be precisely characterized
because the sample did not have asteroseismic evolutionary state
classifications. More fundamentally, the stellar parameters were not
tested against a fundamental scale (interferometric radii, for
example). Theoretically-motivated corrections to $\dnu$ were not
applied to the catalogue, meaning that there were $\approx 10\%$-level
systematic offsets in the RGB mass and radius scales. Indeed,
\cite{epstein+2014} would discover that APOKASC-I asteroseismic radii
and masses were systematically offset compared to the old stellar
population in the halo.

The APOKASC-2 catalogue \citep{pinsonneault+2018} improved upon its
predecessor in these and other ways. The new catalogue was calibrated
to the dynamical mass scale from two clusters, NGC 6791 and NGC
6819. It also contained evolutionary state information, theoretical
$\dnu$ corrections were applied, and a self-consistent asteroseismic
scale and error budget were derived using asteroseismic parameters
from five independent pipelines.

The current work capitalizes on this catalogue to perform a test of
the scaling relations themselves. With the stellar parameters
calibrated to a fundamental scale, we can compare the calibrated radii
from the catalogue to radii from \textit{Gaia}, effectively using each
\textit{Gaia} radius as its own fundamental calibrator. This allows
us, ultimately, to have not two calibrators (the masses of the giant
branches of NGC 6791 and NGC 6819), but thousands --- testing the
scaling relations at every radius, temperature, and metallicity in the
sample; the fact that \textit{Gaia} provides a distance to each star
means that every star, in effect, is like an open cluster member. Knowing the
distance, in combination with flux, means that one knows the
luminosity, and thus, in combination with a temperature and the
Stefan-Boltzmann law, the radius. This exercise therefore requires
accurate and precise luminosities and temperatures that are not
subject to systematic biases. In what follows, we take care to ensure
that our luminosities and temperatures are well-characterized.

In previous work, 
\cite{huber+2017} applied this technique using {\it Gaia} Data Release
(DR) 1, and demonstrated
that the Tycho-{\it Gaia} astrometric solution (TGAS)
  \citep{michalik_lindegren&hobbs2015,gaia2016} and asteroseismic radii agreed to within $5\%$ for
  stars with radii of $\approx 0.8 - 8 \rsun$. A similar exercise was also
performed with {\it Hipparcos} \citep{vanleeuwen2007} parallaxes \citep{silva_aguirre+2012},
indicating agreement at the $5\%$ level. More recently,
\cite{sahlholdt+2018b} used \textit{Gaia} DR2 parallaxes to test the
dwarf asteroseismic radius scale, finding that it is concordant with
\textit{Gaia} radii at the 2-3$\%$ level. The red clump radius scale
has also been shown to agree with the {\it Gaia} radius scale at the
$2\%$ level \citep{hall+2019}. Most recently, a determination of the
{\it Gaia} parallax zero-point by \cite{khan+2019} 2019 suggests good
agreement between asteroseismic parallaxes and {\it Gaia} DR2 parallaxes
among both first-ascent red giant branch and red clump stars.

The scaling relation radius
scale has been tested in other work against other fundamental
scales, which have all indicated that the asteroseismic radius scale is good to at least the $10\%$ level. Asteroseismic radii have been tested against interferometric
values \citep{huber+2012}, for instance, demonstrating good
agreement. There are a handful of studies comparing the asteroseismic scale to a
dynamical scale using eclipsing binaries. Following studies of individual binary systems hosting a giant star by \cite{frandsen+2013}
and \cite{rawls+2016a}, \cite{gaulme+2016} contributed the largest such analysis. All of the red giants from \cite{gaulme+2016} have dynamical and asteroseismic
radii less than $15\rsun$, and exhibit an offset at the $5\%$ level in
the sense that the asteroseismic radii are larger than the dynamical
radii.  \cite{brogaard+2018}, however, using a subset of the \cite{gaulme+2016} sample, argued that a reanalysis of the
stellar parameters brought the asteroseismic radii into agreement
with the dynamical radii.

This paper models itself after \cite{huber+2017}, improving upon those
constraints thanks to the increased precision of {\it Gaia} DR2 \citep{gaia2016,gaia2018} parallaxes over those from DR1. We also expand the analysis to include stars with a radius
of up to $\sim 50\rsun$. Here, we look at $4128$ stars with asteroseismic radii and parallaxes from {\it Gaia} DR2,
comprising $372$ dwarfs and $3755$ giants. Note that we are analyzing 
first-ascent RGB stars only; thus our giant sample is a subset of the
nearly 7000 stars of APOKASC-2. Given that there
are known red clump versus RGB systematics, we analyze red clump stars
separately (Pinsonneault et al., in prep.).

  A comparison of the {\it Gaia} DR2 radius scale and the asteroseismic radius
  scale will be sensitive to all of
  the scales involved: the luminosity scale (which depends on the {\it Gaia} parallax scale and the bolometric
  correction scale), the temperature scale, and the asteroseismic
  radius scale. In this work, we
  use {\it Gaia} parallaxes corrected according to \cite{zinn+2019} as a benchmark against which to compare the asteroseismic radius scale. We also quantify the systematic errors in the bolometric correction scale and the temperature scale by comparing to other scales established in the literature. We also quantify the spatial correlations in {\it Gaia} DR2
  parallaxes for the \textit{Kepler} field, following the example
  of \cite{zinn+2017a}. Such correlations are directly relevant to other population-level studies, which compute some sky-averaged statistic that combine quantities that depend on parallax (e.g., open cluster distance calculations).

\section{Data}
\label{sec:data}
\cite{zinn+2019} presented the basic {\it Gaia}-asteroseismic data set
we use in this paper, and we review its properties here.

\subsection{The asteroseismic comparison samples}

As mentioned in \S\ref{sec:intro}, asteroseismology offers
so-called scaling relations, which are means of deriving stellar
masses and radii based on the characteristic frequencies of solar-like oscillations, $\dnu$
and $\numax$. The radius scaling relation is the subject of study in
this work, and takes the form

\begin{equation}
\frac{R}{\rsun} \approx \left( \frac{\numax}{f_{\numax} \numaxsun}\right)  \left( \frac{\dnu}{f_{\dnu} \dnusun}\right)^{-2} \left(\frac{\teff}{\teffsun}\right)^{1/2}.
\label{eq:scaling3}
\end{equation}

This relation bears the qualification ``scaling'' because it re-scales
the solar values of $\rsun$, $\numaxsun$, $\dnusun$, and $\teffsun$
based on relations between 1) $\dnu$ and the density of a
star \citep{tassoul1980,christensendalsgaard1993}, and 2) $\numax$ and the surface gravity \&
temperature of a star \citep{kjeldsen&bedding1995,brown+1991}, formalized in their own scaling relations as follows:

\begin{equation}
\label{eq:numaxscaling}       
\frac{\dnu}{f_{\dnu} \dnusun} \approx \sqrt{\frac{M/\msun}{(R/\rsun)^3}}                             
\end{equation}
and
\begin{equation}
\label{eq:dnuscaling}    
\frac{\numax}{f_{\numax} \numaxsun} \approx \frac{M/\msun}{(R/\rsun)^2\sqrt{(\teff/\teffsun)}}
\end{equation}

We use
the same solar values for these quantities as used in constructing the
APOKASC-2 catalogue \citep{pinsonneault+2018}: $\numaxsun =
3076 \muhz$, $\dnusun = 135.146 \muhz$, and $\teffsun = 5772 K$.

Theoretically-motivated corrections to observed $\dnu$, denoted in the above equations as $f_{\dnu}$, are required to bring the observed $\dnu$ into agreement with the
theoretical $\dnu$ assumed in asymptotic pulsation theory. These corrections
depend on the evolutionary state of the star, as well as the
mass, temperature, surface gravity, and metallicity \citep[e.g.,][]{sharma+2016}. Similar corrections may be required of $\numax$ (denoted $f_{\numax}$ in the above equations), and, if present and not accounted for, would be a potential source of problems in the asteroseismic radius scale. Throughout the work, we assume $f_{\numax} = 1$. We discuss the possibility that $f_{\numax}$ departs from unity in a way that depends on metallicity in \S\ref{sec:met}. 

Using the asteroseismic radius scaling relation (Equation~\ref{eq:scaling3}), we derive radii, which we compare to {\it Gaia} radii. For the purposes of this work, we correct the asteroseismic radii using $f_{\dnu}$ given their solid theoretical and empirical basis \citep[e.g.,][]{white+2011,sharma+2016,guggenberger+2016}, and attempt to interpret remaining discrepancies in the asteroseismic radius scale in terms of proposed $\numax$ corrections, $f_{\numax}$. We test the radius scaling relation in four radius regimes: for the three largest radius regimes, we use a sample consisting of first-ascent RGB stars, and for the smallest radius regime, we use a sample consisting of dwarfs and subgiants. We describe these samples next.

\subsubsection{Giants}
The primary asteroseismic comparison sample in our study is one of $\approx$\,3800 RGB stars from the APOKASC-2
catalogue \citep{pinsonneault+2018}, which have $\numax$ and $\dnu$ values that are averaged across five independent asteroseismology pipelines. Asteroseismic evolutionary state classifications are derived from asteroseismology for all but $\approx$\,200 of these stars, with the remaining categorized as RGB stars based on spectroscopy (see \citealt{holtzman+2018} for a description of the spectroscopic method). The value for $\numaxsun$ from \cite{pinsonneault+2018}, which we also use in this work, was chosen to bring the mean asteroseismic mass into agreement with the dynamical masses of NGC 6791 and NGC 6819. A systematic error on the APOKASC-2 radii of $0.7\%$ is thus inherited from the uncertainty on the open cluster dynamical masses. Temperatures for the radius scaling relation are taken from APOGEE DR14 \citep{holtzman+2018}, as are metallicities for the purposes of computing theoretical $f_{\dnu}$ values. We have adopted theoretical $f_{\dnu}$ from \cite{pinsonneault+2018}, which are computed using a
revised version of the Bellaterra Stellar Parameters Pipeline
\citep[BeSPP][]{serenelli+2013a,serenelli17}. Where noted, we have validated our results using an alternate $f_{\dnu}$ prescription from \cite{sharma+2016}. Our giants have asteroseismic radii greater than $3.5\rsun$.

\subsubsection{Dwarfs and subgiants}
The other asteroseismic comparison sample consists of $\approx$\,400 dwarfs and subgiants
with asteroseismic parameters taken from \cite{huber+2017}, which includes stars from a reanalysis of the
\citet{chaplin14} sample by \citet{serenelli17}, as well as stars from \cite{huber+2013}. As for the giants, effective
temperatures and metallicities are taken from APOGEE DR14, and BeSPP $f_{\dnu}$ are used. We only consider stars with radii less than $3.5\rsun$ from this sample.\footnote{One star present in both the \cite{serenelli17} sample and our giant sample, KIC 10394814, was excluded from the dwarf/subgiant sample.}

The giant $\numax$ and $\dnu$
values in the APOKASC-2 catalogue are on the mean asteroseismic scale,
whereas those for our dwarfs and subgiants are natively on the SYD pipeline
scale \citep{huber+2009}. We correct the asteroseismic parameters to
bring them into alignment with the APOKASC-2 mean scale, which amounts to a negligible re-scaling of $\numax$ and $\dnu$ by $0.06\%$ and $0.05\%$.
Considering we use BeSPP theoretical $f_{\dnu}$ for both the giant and the dwarf/subgiant samples, the end result is that the $\numax$ and $\dnu$ 
values in our full sample spanning dwarfs and giants are on a
consistent system. 

\subsection{The {\it Gaia} Data Release 2 sample}
\label{sec:gaia}
Stellar parallax, $\varpigaia$, constitutes the most important
information from {\it Gaia}, which we use in combination with
APOKASC-2 photometric information to derive radii
against which we test the asteroseismic radius
scale.

The {\it Gaia} DR2 parallaxes are of excellent quality, with typical
statistical errors of $0.05$mas for the sort of bright stars that are
in our sample. Some parallaxes, however, may be erroneous due to
unresolved binary motions or statistical errors in the {\it Gaia}
red and/or blue passband. We therefore apply quality cuts to the {\it Gaia} data according
to \cite{lindegren+2018}, by only selecting stars that fulfill the
following criteria, which are the same as used in \cite{zinn+2019}.

\begin{enumerate}
  \item \texttt{astrometric\_excess\_noise} = 0\,;
   \item $\chi \equiv
    \sqrt{\chi^2/n}$, $\chi < 1.2\mathrm{max}(1, \exp{-0.2(G -
      19.5)})$\,;
  \item \texttt{visibility\_periods\_used} > 8\,;
  \item $1.0 + 0.015(G_{BP} - G_{RP})^2 <$
    \texttt{phot\_bp\_rp\_excess\_factor} $< 1.3 + 0.06(G_{BP} - G_{RP})^2$\,;
\end{enumerate}
where $\chi^2 \equiv$ \texttt{astrometric\_chi2\_al}, $n \equiv$
\texttt{astrometric\_n\_good\_obs\_al} - 5, $G_{BP} =
\texttt{phot\_bp\_mean\_mag}$, $G_{RP} = \texttt{phot\_rp\_mean\_mag}$, $G = \texttt{phot\_g\_mean\_mag}$.

The first and second cuts remove stars with a bad parallax solution, which may be
caused by unresolved binary motion. The third cut rejects stars whose
{\it Gaia} observations are over time baselines that are not
well-separated, and therefore whose underlying astrometric data does
not constrain the astrometric model very well. The fourth cut removes stars that
are plagued by bad {\it Gaia} photometry. 43 stars were rejected by these cuts for the dwarf/subgiant sample, and 182 from the giant sample.

We apply a final quality cut to remove stars whose asteroseismic
parallaxes (which are derived according to the next section) and {\it
Gaia} parallaxes do not agree at the $5\sigma$ level. This cut is
performed for each analysis method described in \S\ref{sec:methods}. One star from the dwarf/subgiant sample are rejected in this way, and 15 from the giant sample.

Photometric information and temperatures are required to compute a radius from a parallax and vice-versa, as discussed in the next section. We adopt Two Micron All Sky Survey
\citep[2MASS;][]{skrutskie+2006} $\ks$ photometry, rejecting 11 RGB stars without reliable photometric uncertainty (photometric quality flag of 'F'). We use APOGEE DR14 temperatures to perform these transformations. For the giants in our analysis, \cite{rodrigues+2014} extinctions from the APOKASC-2 catalogue are used to apply small de-extinction corrections to the infrared photometry. For
the $\ks$ extinction coefficient, we use the \cite{fitzpatrick1999}
reddening law applied to the 2MASS $\ks$ passband, as implemented in
\texttt{mwdust} \citep{bovy+2016}, assuming a $E(B-V)$
from \cite{sfd1998}, as re-calibrated by \cite{sf2011}. The dwarf and subgiant extinction values are from \cite{green+2015}.

Our final sample consists of 328 dwarfs/subgiants and 3554 RGB stars.

\section{Methods}
\label{sec:methods}
The naive approach to testing the asteroseismic radius scaling relation would be to compare APOKASC-2 asteroseismic radii to the radii released as part of {\it Gaia} DR2. However, the out-of-the-box {\it Gaia} DR2 radii were derived without modeling extinctions, without correcting for the known DR2 parallax zero-point errors, and with temperatures that are not on the same scale as the APOGEE DR14 temperatures used to compute our asteroseismic radii. Therefore, we compute our own set of radii using the {\it Gaia} DR2 parallaxes, and adopt temperatures and extinctions from APOKASC-2. To do this, we use the Stefan-Boltzmann law to invert a luminosity (from an observed flux and bolometric correction in combination with a {\it Gaia} DR2 distance) plus a temperature to yield a radius.

The {\it Gaia}-asteroseismology radius comparison requires not only a temperature, extinction, bolometric correction, and a scaling relation radius, but also a {\it Gaia} parallax, of course. The {\it Gaia} parallaxes suffer from a small but non-negligible zero-point offset that is position-dependent and appears to be dependent on color and magnitude, as well. This needs to be taken into account. Fortunately, our dataset spans a range in both radius and parallax/distance. That means, for a given radius, there are stars that are very close by and stars that are far away. One the one hand, the nearby stars have relatively large parallax, and therefore their {\it Gaia} radii are not sensitive to a relatively small zero-point correction. On the other hand, the distant stars have a relatively small parallax, and their radii are sensitive to zero-point corrections. We use the range in distance in our sample to our advantage by applying our primary analysis to a sub-sample of our asteroseismic comparison sample consisting of stars with large parallaxes whose {\it Gaia} radii are therefore not sensitive to {\it Gaia} parallax zero-point errors.  As we describe in the next section, we fit for radius correction factors among this sub-sample that bring the asteroseismic radius scale in agreement with the {\it Gaia} radius scale, after correcting the {\it Gaia} parallaxes according to \cite{zinn+2019}. In practice, we do this by working in parallax space and not radius space: we use the Stefan-Boltzmann law to transform our asteroseismic radii, in combination with fluxes and temperatures, into distances/parallaxes. As we note in \S\ref{sec:abs}, the asteroseismic parallax is more sensitive to problems in the asteroseismic radius scale for large parallax stars than small parallax stars, which is another benefit of applying our primary analysis to large parallax stars. The rest of the stars with smaller parallaxes are then used to further validate the differential trends we see in the radius agreement as a function of evolutionary state (\S\ref{sec:diff2}), and to validate the choice in our {\it Gaia} parallax zero-point correction (\S\ref{sec:choices}).

Elements of this approach are described in \cite{zinn+2019}, wherein the authors derived a {\it Gaia} DR2 parallax zero-point for the {\it Kepler} field assuming the asteroseismic radii were not subject to errors. This assumption is valid given the relative insensitivity of the inferred parallax offset to the asteroseismic radius scale (see their Figure 5b). We discuss this assumption further in \S\ref{sec:choices}, and demonstrate that the {\it Gaia} DR2 parallax zero-point we adopt does not bias our results. Ultimately, we use the \cite{zinn+2019} {\it Gaia} DR2 parallax zero-point to correct the {\it Gaia} parallaxes and derive {\it Gaia} radii, against which we compare the asteroseismic radius scale.

To test the asteroseismic radius scale, we begin by constructing an asteroseismic parallax, $\varpiseis$, based on an effective temperature, $\teff$, and bolometric flux, $F$:

\begin{align}
  \label{eq:plx}
  \varpiseis(\teff, F, \invrscal) &= F^{1/2} \sigma_{\mathrm{SB}}^{-1/2} \teff^{-2}
  \invrscal \\
  &= f_0^{1/2}10^{-1/5(m + BC - A_m)} \sigma_{\mathrm{SB}}^{-1/2} \teff^{-2} \invrscal \notag,
\end{align}
where the bolometric flux is computed based on a magnitude, $m$, a
bolometric correction for that band, $BC$, a flux zero-point
calibrated for that band, $f_0$, and an extinction in that
band, $A_m$. $\sigma_{\mathrm{SB}}$ is the Stefan-Boltzmann constant,
and the stellar radius, $R$, is taken to be the asteroseismic radius, $\rscal$,
which is derived from the radius scaling relation (Equation~\ref{eq:scaling3}).

Like the approach from \cite{zinn+2019}, we then model the differences in asteroseismic and {\it Gaia} parallaxes. In that work, the authors fit a three-parameter model that described a global, color- and magnitude-dependent parallax zero-point such that the asteroseismic parallaxes and {\it Gaia} parallaxes agreed. In this work, we adopt the zero-point from \cite{zinn+2019}, and then fit for asteroseismic radius correction factors that minimize the difference between the two parallax scales. We describe this model in the next section.

\subsection{Scaling radius correction model}
We are interested in comparing asteroseismic radii to those derived using classical constraints from a combination of $L$ and $\teff$.  As there are physical effects that could be radius-dependent, we begin by defining distinct radius regimes where we will test our agreement. We can therefore test not only for problems in the radius scaling relation, but also whether the asteroseismic-{\it Gaia} radius agreement is different for evolved stars in different radius regimes. The smallest radius regime that we explore is the dwarf/subgiant regime, with radii less than $3.5\rsun$, and down to $\approx 0.8\rsun$. The other radius regimes we consider are all stages on the first-ascent RGB. The low-luminosity RGB stars below the radius of the red clump, $3.5\rsun \geq R \leq 10 \rsun$ and more evolved RGB stars with $10\rsun < R < 30 \rsun$ comprise the next two radius regimes. The largest radii that we consider in our analysis are those for which $R \geq 30\rsun$.

In order to identify problems in the asteroseismic radius scale, we fit for an asteroseismic radius correction factor in each of the above radius regimes. We do so after correcting for the \textit{Gaia} parallax zero-point described by a global offset, $c = 52.8 \muas$; an astrometric pseudo-color ($\nueff$)-dependent offset, $d = -151.0 \muas \mu m$; and a {\it Gaia} $G$-band magnitude-dependent offset, $e = -4.20 \muas/\mathrm{mag}$ \citep{zinn+2019}. We fit for the
radius anomalies, $a_1$, $a_2$, $a_3$, and $a_4$, such that they minimize the difference
between $\hatvarpiast$ and $\hatvarpigaia$.  In parallax space, this is written as:

\begin{equation}
\hatvarpigaia = \\
\begin{cases}
a_1\hatvarpiast - z & R < 3.5 \rsun \\
a_2\hatvarpiast  - z & 3.5 \rsun \geq R \leq 10 \rsun \\
a_3\hatvarpiast  - z &
10\rsun < R < 30 \rsun \\
a_4\hatvarpiast  - z & R \geq 30\rsun\,, \\
\end{cases}
\label{eq:broken}               
\end{equation}
where $z$ describes the {\it Gaia} parallax zero-point correction:
\begin{equation}
\label{eq:zp}
z =    c +  d(\hatnueff - 1.5) + e(\hat{G} - 12.2)\,.
\end{equation}

We turn our model for $\hatvarpiast - \hatvarpigaia$ into a likelihood by assuming Gaussian errors and a covariance matrix describing
the covariance in parallax space of two stars, $i$ and $j$
separated by an angular distance, $\Delta \theta_{ij}$, which
reads

\begin{equation}
C_{ij}(\Delta \theta_{ij}) = \chi(\Delta \theta_{ij})\sigma_{\varpigaiai}\sigma_{\varpigaiaj} + \delta_{ij}\sigma^2_{i},
\label{eq:cov}
\end{equation}
where $\chi(\Delta \theta_{ij})$ is the spatial correlation in the
parallaxes of the stars (see Appendix~\ref{app:cov}); $\sigma_{\varpigaiai}$ is the
{\it Gaia} parallax error for star $i$; $\sigma_i$ is the uncertainty on $\hatvarpiasti - \hatvarpigaiai$; and $\delta_{ij}$ is the Kronecker
delta function. Hence, for $i = j$, $C_{ij}(\Delta \theta_{ij} = 0) =
\sigma^2_{i} = \sigma^{2}_{\varpigaiai} + \sigma^2_{\varpiasti}$. We defer a discussion of the off-diagonal elements of $C$ to Appendix~\ref{app:cov}, and report our radius agreement result (\S\ref{sec:abs_results}) with and without spatial parallax correlation terms in $C$. Our results are unaffected by the level of spatial correlation present in the high-parallax sub-sample due to the sparsity of these stars in the {\it Kepler} field. If we were making inferences using the full sample of $\sim 3900$ stars, these spatial correlations would inflate uncertainties in averaged values at the $10\%$ level.

We therefore write the likelihood for the parameters of interest, $a_1$, $a_2$, $a_3$, and $a_4$, as:

\begin{align}
  \label{eq:second_model}
  \begin{split}
&\mathcal{L}(a_1,a_2,a_3,a_4| c, d, e, \hatvarpigaia, \hatteff,
    \hatdnu, \hatnumax, \hat{A}_V,\\
    &\hat{\ks}, \hat{BC}, \hat{G}, \hatnueff) \propto \\
&\frac{1}{\sqrt{(2 \pi)^N \mathrm{det}\, C}} \exp{\left[-\frac{1}{2} (\vec{y} -
  \vec{x})^{\mathrm{T}}  C^{-1} (\vec{y} - \vec{x})\right]},
\end{split}
\end{align}
where
\begin{equation*}
  \vec{y} \equiv \\
  \begin{cases}
 a_1\hatvarpiast(\hatteff, \hatdnu,
      \hatnumax, \hat{A}_V, \hat{\ks}, \hat{BC}) & R < 3.5 \rsun \\
 a_2\hatvarpiast(\hatteff, \hatdnu,
      \hatnumax, \hat{A}_V, \hat{\ks}, \hat{BC}) & 3.5 \rsun \geq R \leq 10 \rsun \\
       a_3\hatvarpiast(\hatteff, \hatdnu,
      \hatnumax, \hat{A}_V, \hat{\ks}, \hat{BC}) & 10\rsun < R < 30  \rsun \\
      a_4\hatvarpiast(\hatteff, \hatdnu,
      \hatnumax, \hat{A}_V, \hat{\ks}, \hat{BC}) & R \geq 30\rsun \\
  \end{cases}
\end{equation*}
and
\begin{equation*}
\vec{x} \equiv \hatvarpigaia  +  c + d(\hatnueff - 1.5) + e(\hat{G} - 12.2).
\end{equation*}

The only free parameters in our asteroseismic radius correction model are $a_1$, $a_2$, $a_3$, and $a_4$ because {$c$, $d$, $e$} are fixed to the values from \cite{zinn+2019}. We fit for the mean values and uncertainties in $\{a_1, a_2, a_3, a_4\}$ with
MCMC, as implemented with the \texttt{emcee} package \citep{foreman-mackey+2013}. To do
so, we work with the posterior probability for $\{a_1, a_2, a_3, a_4\}$, which
is the likelihood multiplied by any priors we may have on the
parameters. We apply the priors that the radius correction factors should not be larger
than $1.2$ or less than $0.8$, which is borne out by previous studies that find problems in the radius scaling relations appear to be at less than the $5\%$ level \citep{gaulme+2016,brogaard+2018,huber+2017,sahlholdt+2018b}.

In this work, we adopt an infrared bolometric
correction. This choice means that the bolometric correction is much
less dependent on temperature because the $K_{\mathrm{s}}$-band is
only linearly sensitive to temperature for a blackbody with the
temperature of a cool giant (instead of
exponentially sensitive in the visual band). Effects due to dust absorption are also
markedly reduced in the infrared compared to the visual. The bolometric correction is interpolated
from MIST bolometric correction tables \citep{dotter+2016a,choi+2016a,paxton+2011a,paxton+2013a,paxton+2015a}, which are computed from the C3K grid of 1D atmosphere models (Conroy
et al., in prep; based on ATLAS12/SYNTHE;
\citealt{kurucz+1970a,kurucz+1993a}). We discuss the effects of our
choice of bolometric correction in \S\ref{sec:bc_syst}.

\subsection{Systematics due to the luminosity scale}
\label{sec:bc_syst}
The luminosities that enter into our radius comparison have two components that admit systematic uncertainties: the bolometric flux scale and the parallax scale.

The parallax systematic is easily understood to be an additive systematic, since our radius comparison is performed by converting asteroseismic radii into parallaxes (Equation~\ref{eq:broken}). By adopting the {\it Gaia} parallax zero-point from \cite{zinn+2019}, we admit a systematic uncertainty of $8.6\muas$ in our parallax difference comparison (Equation~\ref{eq:broken}) due to the uncertainty on $c$ (Equation~\ref{eq:zp}). This corresponds to a $\approx 1.3\%$ systematic in radius space for a typical giant in our sample, and even less among our dwarfs and subgiants because they have larger parallaxes.

Systematics in the bolometric correction and extinction scales enter into our analysis when converting an asteroseismic radius into an asteroseismic parallax via the flux term, $F$, in Equation~\ref{eq:plx}. This means that a systematic in the bolometric correction or extinction of $X$ mag introduces a $\frac{X}{2}\%$ systematic in our radius comparison. We explore the sensitivity of our reported giant radius correction factors on the choice of bolometric correction and extinction by using an alternate extinction scale and five alternate bolometric corrections. 

The extinction scale is tested using a spectral energy distribution (SED) approach, and it also provides an independent check on the bolometric correction.  With the SED method, a bolometric correction
is not required because the entire SED is fitted, and extinction is
computed simultaneously, based on the SED shape. This process is described
in \cite{stassun+2016a} and \cite{stassun+2017a}. We have also tested the robustness of our results by using the \cite{ghb09}
InfraRed Flux Method (IRFM) bolometric flux scale; the \cite{ghb09} $\ks$-band bolometric flux scale; the MIST $g$-band bolometric flux scale; and the \cite{flower+1996} $V$-band bolometric flux scale. More details on these checks of bolometric correction and extinction systematics are found in Appendix~\ref{app:fbol}.

Between the self-consistency of the MIST bolometric corrections and comparisons to
independent systems described further in Appendix~\ref{app:fbol}, we conclude that the $\ks$-band  bolometric correction may have a systematic
error of up to $1.9\%$, meaning the radii are good to at least $1.0\%$, which we take as a systematic error due to bolometric correction and extinction choice.

\subsection{Systematics due to the temperature scale}
\label{sec:teff_syst}
Our radius comparison is more sensitive to temperature scale systematics than the above luminosity systematics because $R_{\mathrm{seis}}/R_{Gaia} \propto T^{5/2}$ as opposed to $R_{\mathrm{seis}}/R_{Gaia} \propto L^{-1/2}$ (see Equations~\ref{eq:scaling3}~\&~\ref{eq:plx}). The APOGEE DR14 temperatures we adopt for both giants and dwarf/subgiants have been calibrated to be on the \cite{ghb09} IRFM scale. Therefore, the predominant systematic possible in the temperature scale used in this work is the systematic in the fundamental IRFM scale. Work on the IRFM scale dates back decades \citep{blackwell+1977a,blackwell+1980a}, and has had widespread application in astronomy due to its relative insensitivity to metallicity, surface gravity, and model atmospheres \citep[e.g.,][]{arribas+1987a,alonso+1994a}. Recently, \cite{casagrande+2010} determined that the IRFM scale for dwarfs and subgiants is good to at least $30$-$40$K when comparing to other temperature scales. They concluded that any small temperature systematics that may exist in the IRFM scale are likely due to the underlying accuracy of infrared photometric calibrations and Vega zero-points. Similarly, in the giant regime, \citep{ghb09} found that their IRFM implementation agreed to within $\approx 40$K with the prevailing giant IRFM temperature application in the literature \citep{alonso+1999a}, for the metallicity
range of the majority of stars considered in this work
($-0.4 $< [Fe/H] < $0.4$). These systematics, when taken to be $2\sigma$ errors, imply that there is a systematic uncertainty in the radius scale due to the temperature scale used in this work of up to $1.1\%$ at the $1\sigma$ level. Because the APOGEE temperatures are adjusted to be on a fundamental scale, any inferred temperature difference must therefore be in the fundamental system, not on uncalibrated spectroscopic measurements that have much larger systematics (see \cite{casagrande+2010} for an extensive discussion).

\subsection{Systematics due to the asteroseismic radius scale}
\label{sec:rad_syst}
Note that due to the calibration of the APOKASC-2 asteroseismic data to open cluster dynamical masses, the asteroseismic radii for giants and dwarfs/subgiants port over a systematic uncertainty of $0.7\%$ from the dynamical mass scale random uncertainty. This means that when we go on to test the asteroseismic radius scale, all the reported agreements have an implicit systematic uncertainty of $0.7\%$.

\subsection{Total systematic uncertainty in radius comparison}
Adding in quadrature the systematic uncertainties from \S\ref{sec:bc_syst}-\S\ref{sec:rad_syst}, we estimate a total systematic uncertainty of $2.0\%$ in our {\it Gaia}-asteroseismology radius scale comparison.

\subsection{A sub-sample for determining the absolute accuracy of the scaling relations}
\label{sec:abs}
\begin{figure}
\includegraphics[width=0.5\textwidth]{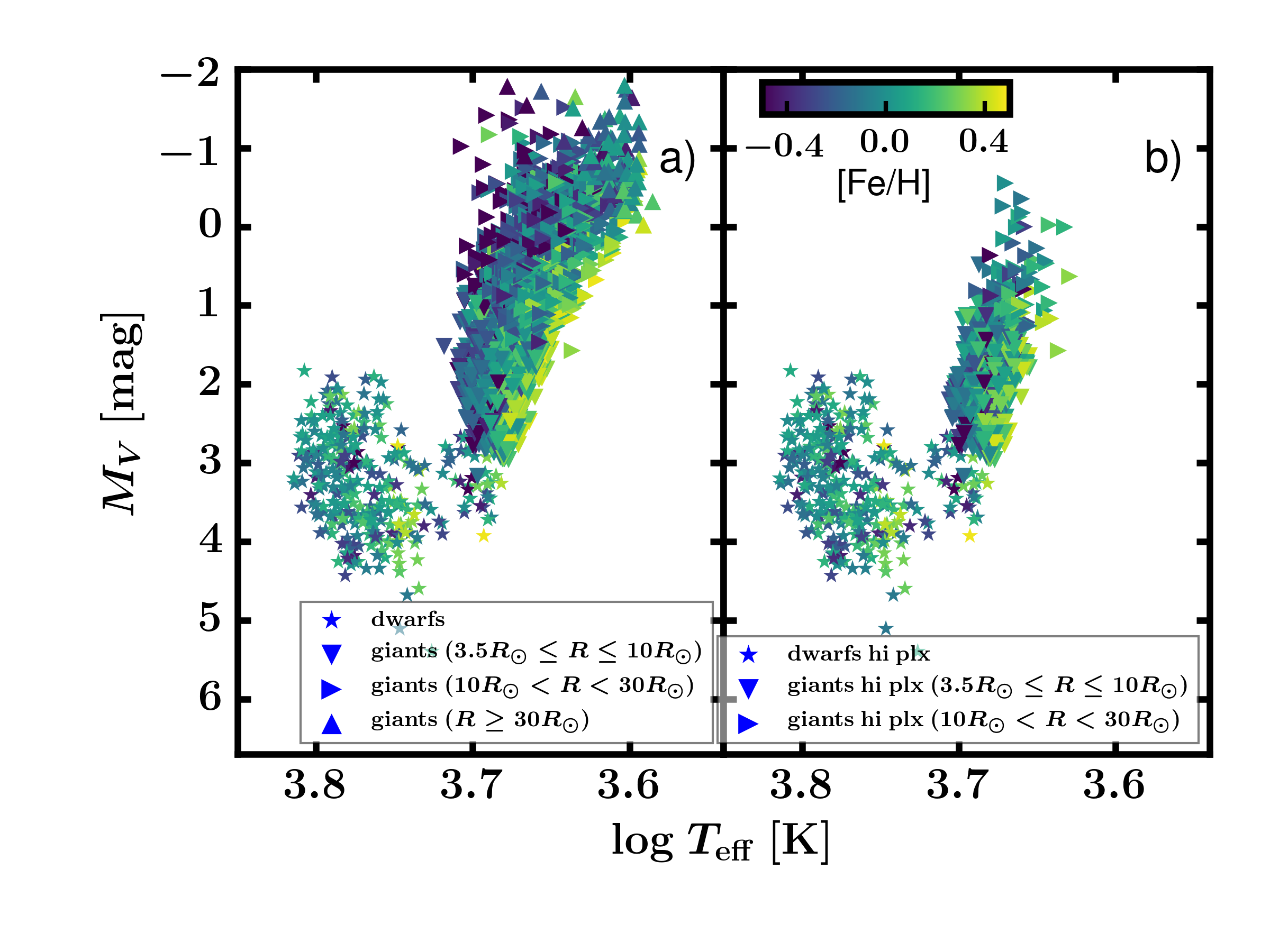}
\caption{HR diagram showing the full giant \& dwarf/subgiant samples (left) and the high-parallax sub-sample (right) used in this work, divided into the four different radius regimes we consider.}
\label{fig:hr}
\end{figure}

\begin{figure}
\includegraphics[width=0.5\textwidth]{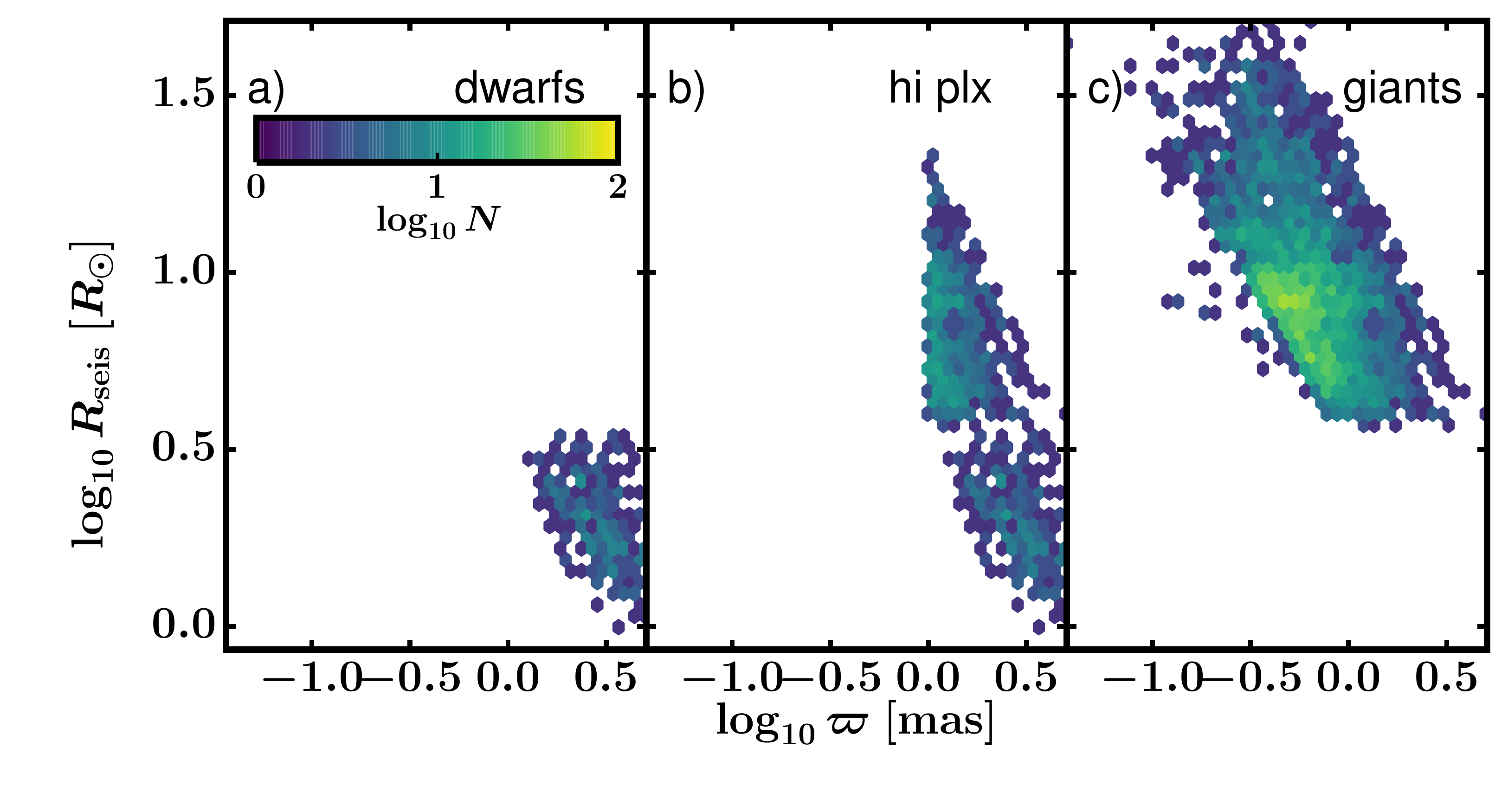}
\caption{The distribution in parallax-radius space of the dwarf sample (left), the high-parallax sub-sample (middle), and the full giant sample (middle) used in this work.}
\label{fig:radplx}
\end{figure}

The primary goal of this work is to test the accuracy of the radius scaling
relation. To do so, we need to ensure that the {\it Gaia} parallaxes
themselves are on an absolute scale.
\cite{zinn+2019} have looked at the issue of zero-point errors in {\it
  Gaia} parallaxes by assuming that the asteroseismic parallaxes were
on an absolute scale and correcting the {\it Gaia} parallaxes to minimize the difference between the two scales. They showed that asteroseismic radius problems of the sort we are looking for in this work would manifest as a difference in {\it Gaia} and asteroseismic parallax scales that is larger at larger parallaxes (see their Figure 2). Furthermore, any {\it Gaia} zero-point errors are not as important among high-parallax stars as they are for small-parallax stars (see \S\ref{sec:choices}). For these two reasons, we constructed a
high-parallax sub-sample consisting of stars with $\varpi > 1\mas$, which will be the population from which we
infer our best-fitting model for the asteroseismic
radius correction model (Equation~\ref{eq:broken}).  Its distribution in the HR diagram and in parallax-radius space are shown in Figures~\ref{fig:hr}b~\&~\ref{fig:radplx}b. To compute the absolute magnitudes, we used distances based on {\it Gaia} DR2 parallaxes, calculated following \cite{bailer-jones+2018a}, by using the mode of the likelihood with an exponentially-decreasing volume density prior with scale length 1.35kpc. All of the dwarfs and subgiants are included in this sub-sample, given their relatively close distances. However, none of the stars with $R \geq 30\rsun$ has a parallax that satisfies the $\varpi > 1\mas$ high-parallax sub-sample selection criterion. Therefore, $a_4$ is inferred using all of the stars with $R \geq 30\rsun$, regardless of parallax. As we argue in \S\ref{sec:choices}, it does not appear that $a_4$ should be significantly biased by this choice.

\subsection{A sample for determining differential corrections to the radius scaling relation along the giant branch}
\label{sec:diff}
Whereas we believe the high-parallax sample described in the previous section gives the best estimate of the asteroseismic radius scaling
relation corrections, we can also evaluate the agreement between {\it Gaia} and asteroseismic radius for stars at all parallaxes, and with a larger number of stars than the high-parallax sub-sample. For this purpose, we use all of our giant sample, whose distributions in the HR diagram and in parallax-radius space are shown in Figures~\ref{fig:hr}a~\&~\ref{fig:radplx}c. This sample, which includes small-parallax stars, will also prove useful to demonstrate that {\it Gaia} parallaxes have been adequately corrected for the zero-point offsets (see \S\ref{sec:choices}). 
 
\section{Results}
\label{sec:results}
\subsection{Absolute radius agreement}  
\label{sec:abs_results}
\begin{figure}
\begin{center}
\resizebox{\hsize}{!}{\includegraphics{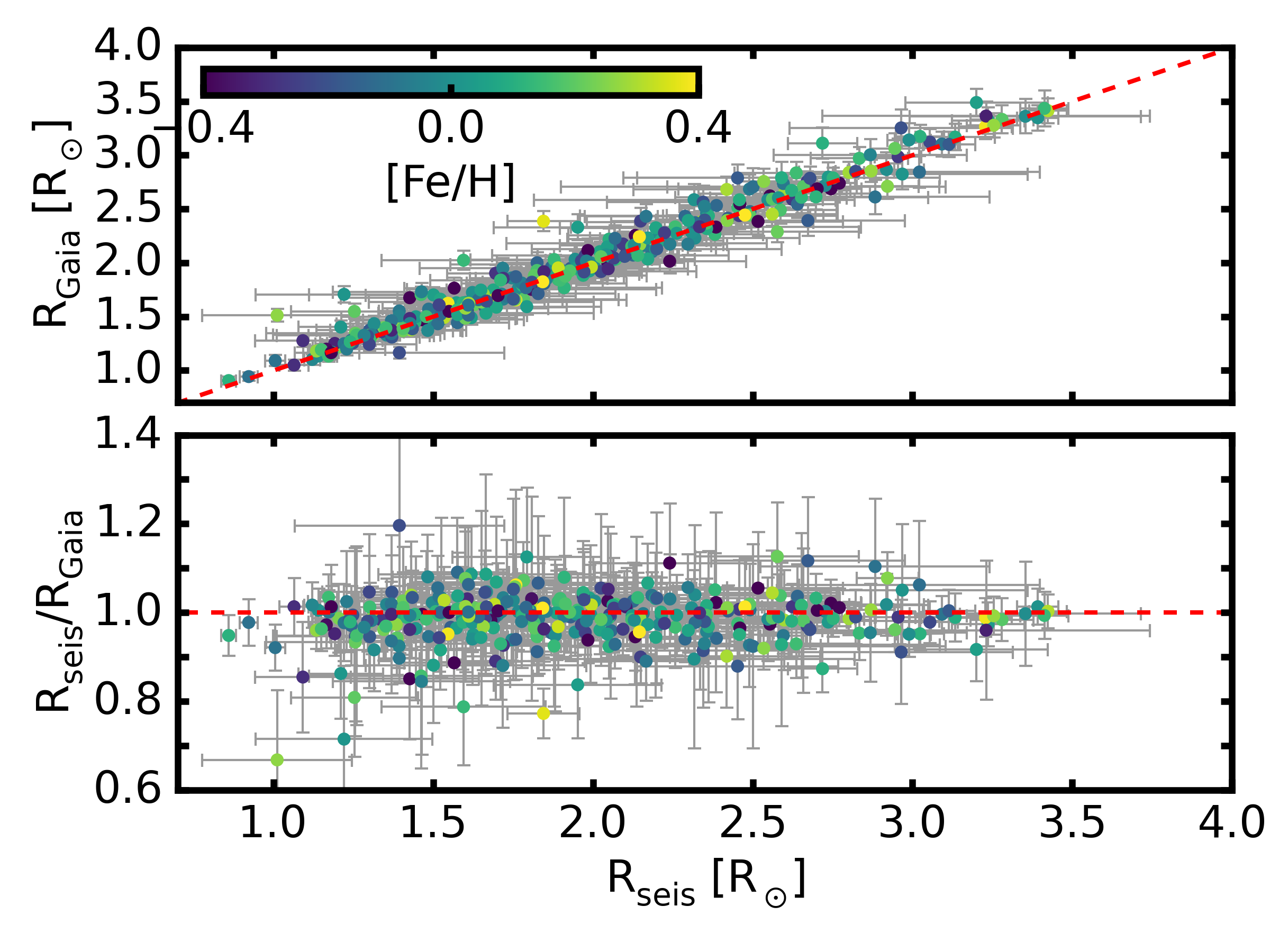}}
\caption{Comparison of radii derived using {\rm Gaia} DR2 parallaxes with radii calculated from asteroseismic scaling relations for the sample in \citet{huber+2017}. Color-coding denotes the metallicity for each star. The average residual median and scatter is $\sim$\,2\% and $\sim$\,5\%, respectively.}
\label{fig1}
\end{center}
\end{figure}

Figure \ref{fig1} compares asteroseismic and {\it Gaia} radii for dwarfs/subgiants, color-coded by metallicity, and plotted without any radius correction factor applied to the asteroseismic radii. The agreement is excellent, with a median offset of $\approx$\,1\% and scatter of $\approx$\,4\%. We observe no strong dependence of the residuals on metallicity, consistent with the results for the larger and more evolved giant sample discussed in \S\ref{sec:met}. The radius correction factor we find in this, the smallest radius regime we consider ($R < 3.5 \rsun$), is $a_1 = 0.979 \pm 0.005 \mathrm{\,(rand.)} \pm 0.020 \mathrm{\,(syst.)}$. This means that the asteroseismic radius scale for dwarfs and subgiants agree with the {\it Gaia} radius scale within the uncertainties.

\begin{figure*}[!p]
\includegraphics[width=0.95\textwidth]{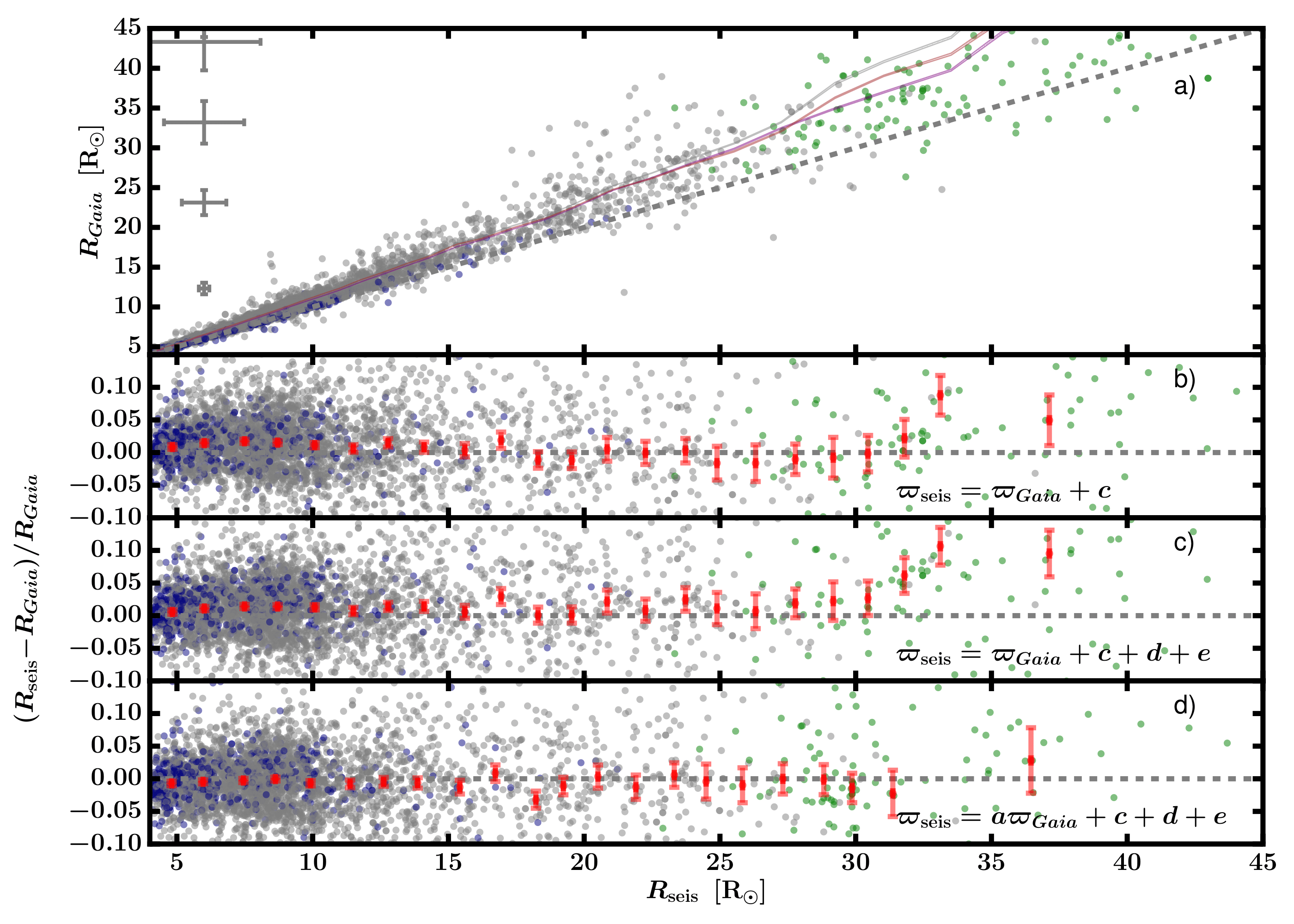}
\caption{Asteroseismic RGB radii are in excellent agreement with {\rm Gaia} radii, which indicates that the asteroseismic radius scaling relation is good to within $2\% \pm 2\%$ up to radii of $30\rsun$. Panel a shows {\rm Gaia} radius as a function of asteroseismic radius for the giants in our sample. Green points are stars with surface gravities, $\log g < 1.6$ ($R \gtrsim 30\rsun$), the regime in which there could be measurement-error related radius systematics \citep{pinsonneault+2018}. Navy points are stars that are part of the sample used to fit radius correction factors for the giants, $a_2$, $a_3$, and $a_4$, which have {\rm Gaia} parallaxes greater than 1mas (``hi plx'' in Figures~\ref{fig:hr} \&~\ref{fig:radplx}). The error bars indicate median errors as a function of {\rm Gaia} radius. Panels b-d show the residuals in the radius agreement after successively correcting the data according to the model of Equation~\ref{eq:broken}, with red error bars showing binned uncertainties on the median: panel b includes a global offset to the {\rm Gaia} parallaxes of $52.8 \muas$ (brown curve in panel a); panel c further includes color- and magnitude-dependent terms of $-151.0 \muas \mu m$ and $-4.20 \muas/\mathrm{mag}$ (grey curve in panel a); panel d finally also corrects the asteroseismic radii by factors $a_2 = 1.015 \pm 0.0025$,  $a_3 = 1.019 \pm 0.0060$, and $a_4 = 1.087 \pm 0.0092$ (purple curve in panel a). }
\label{fig:radb}
\end{figure*}

Figure~\ref{fig:radb}a shows our main result in the giant regime: asteroseismic radii agree
with those from {\it Gaia} within $2.1\% \pm 2.0\% {\rm\, (syst.)}$. Figure~\ref{fig:radb}b indicates the residuals  
when the parallaxes are only corrected
by a zero-point offset ($c$ in Equation~\ref{eq:zp}). Figure~\ref{fig:radb}c shows the agreement after an additional
correction with color- and magnitude-dependent terms ($d$ and $e$ in Equation~\ref{eq:zp}). Finally, Figure~\ref{fig:radb}d shows the agreement after additionally applying the best-fitting radius correction factors from Equation~\ref{eq:broken}. No matter the {\it Gaia} zero-point model, and across a wide range in radius, the agreement between asteroseismic and {\it Gaia} radii is excellent.

Our best-fitting model that we assume in
Figure~\ref{fig:radb}d is fit using the high-parallax sub-sample of
our giants (``K MIST'' in Table~\ref{tab:results}) described
in \S\ref{sec:abs}. The radius correction factors on the RGB of $\{a_2, a_3, a_4\}  = (1.015 \pm 0.003 \mathrm{\,(rand.)} \pm 0.020 \mathrm{\,(syst.)}, 1.019 \pm 0.006 \mathrm{\,(rand.)} \pm 0.020 \mathrm{\,(syst.)}, 1.087 \pm 0.009 \mathrm{\,(rand.)} \pm 0.020 \mathrm{\,(syst.)})$ indicate that the only statistically significant deviation in the asteroseismic radius scale from the {\it Gaia} radius scale is among the most evolved giants.

At radii larger than $30\rsun$, non-adiabatic effects
should begin to manifest in the atmosphere, certainly leading to
breakdowns in the scaling relations \citep{mosser+2013,stello+2014}. $R > 30\rsun$ also roughly corresponds to the same
 gravity regime (log $g$ < 1.6) in which \cite{pinsonneault+2018} found that the
 APOKASC-2 asteroseismic masses were offset from what the giant branch masses should
 be in the clusters NGC 6791 and NGC 6819. 
These evolved stars with $R \geq 30\rsun$ may have a radius scale that is too large compared to the parallactic radius scale: their radius correction factor ($a_4$ in Equation~\ref{eq:broken}) corresponds to a radius inflation of $8.7\% \pm 0.9\% \mathrm{\, (rand.)} \pm 2.0\% {\rm\, (syst.)}$. In this regime, the asteroseismic measurement of $\numax$ in this regime is ill-defined, given the few number of excited modes, and may therefore be systematically biased. Whether due to measurement systematics or due to the physical assumptions in the $\numax$ and $\dnu$ scaling relations themselves no longer being valid (Equations~\ref{eq:dnuscaling} \&~\ref{eq:numaxscaling}), the result is that the radius scaling relation as it is commonly used appears to break down for $R \geq 30\rsun$.

In Table~\ref{tab:results}, we provide $a_2$ and $a_3$ for different choices of
bolometric correction, extinction, and temperature. The agreement of $a_2$ and $a_3$ for these different test cases is generally within the systematic error due to bolometric correction and extinction of $1\%$. We discuss such systematic differences further in our solution in \S\ref{sec:choices}.

\subsection{Differential radius agreement}
\label{sec:diff2}
As we mention in \S\ref{sec:diff}, thanks to the larger number of stars in the full giant sample compared to just the high-parallax giant sub-sample (see Figure~\ref{fig:radb}d grey points versus navy points), the full giant sample gives an indication of differential trends in the asteroseismology-{\it Gaia} radius agreement.

        \begin{longrotatetable}[!p]
  \vspace{1cm}
  \begin{deluxetable*}{ccccccccc}
    \tabletypesize{\small}
    \setlength{\tabcolsep}{0.02in}
          \tablecaption{Parameters for fitted 
            asteroseismic radius scaling relation corrections
\label{tab:results}}
    \tablehead{\colhead{Method} & \colhead{$a_1$} & \colhead{$a_2$} & \colhead{$a_3$} & \colhead{$a_4$} & \colhead{$A_V$ [mag]} & \colhead{$A_{K\mathrm{s}}$ [mag]} & \colhead{$\chi^2/dof$}  & \colhead{$N$}
    }
    \startdata
    K MIST & $0.979 \pm 0.005$ & ... & ... & ... & $0.079$ & $0.009$ & $0.261^{*****}$ & $  328$ \\ \hline
K MIST & ... & $1.015 \pm 0.003$ & $1.019 \pm 0.006$ & ... & $0.104$ & $0.012$ & $0.579^{*****}$ & $  566$ \\ \hline
K MIST  & ... & ... & ... & $1.087 \pm 0.009$ & $0.212$ & $0.024$ & $1.722^{****}$ & $  112$ \\ \hline
K MIST no cov & ... & $1.015 \pm 0.002$ & $1.019 \pm 0.006$ & ...& $0.104$ & $0.012$ & $0.579^{*****}$ & $  566$ \\ \hline
V & ... &$1.001 \pm 0.003$ & $0.992 \pm 0.007$ & ...& $0.103$ & $0.012$ & $0.525^{*****}$ & $  560$ \\ \hline
V MIST &... & $1.017 \pm 0.003$ & $1.017 \pm 0.008$ & ...& $0.103$ & $0.012$ & $0.511^{*****}$ & $  560$ \\ \hline
IRFM & ... &$1.014 \pm 0.002$ & $1.014 \pm 0.006$ & ...& $0.103$ & $0.012$ & $0.601^{*****}$ & $  556$ \\ \hline
 SED & ... & $0.996 \pm 0.002$ & $0.998 \pm 0.007$ & ... & $0.104$ & $0.012$ & $0.690^{*****}$ & $  531$ \\ \hline
   \enddata
   \tablecomments{The best-fitting parameters $a_1$, $a_2$, $a_3$, and
     $a_4$ for Equation~\ref{eq:broken}. Different choices of
     bolometric correction, extinction, temperature, and spatial
     correlation are considered for fitting $a_2$ and $a_3$. Asterisks denote the level of
  discrepancy with the expected $\chi^2$ given the degrees of freedom; each asterisk (up to and including five) denotes one $\sigma$ in the significance
  of the discrepancy. Also noted are the median extinctions in the
  $V$ and $\ks$-band. Our preferred results are from the ``K MIST'' case, as discussed in the text. All the solutions take into account spatial correlations in Gaia
   DR2 parallaxes except the ``K MIST no cov" case. See text for details.}
  \end{deluxetable*}
\end{longrotatetable}

\begin{figure}
\includegraphics[width=0.49\textwidth]{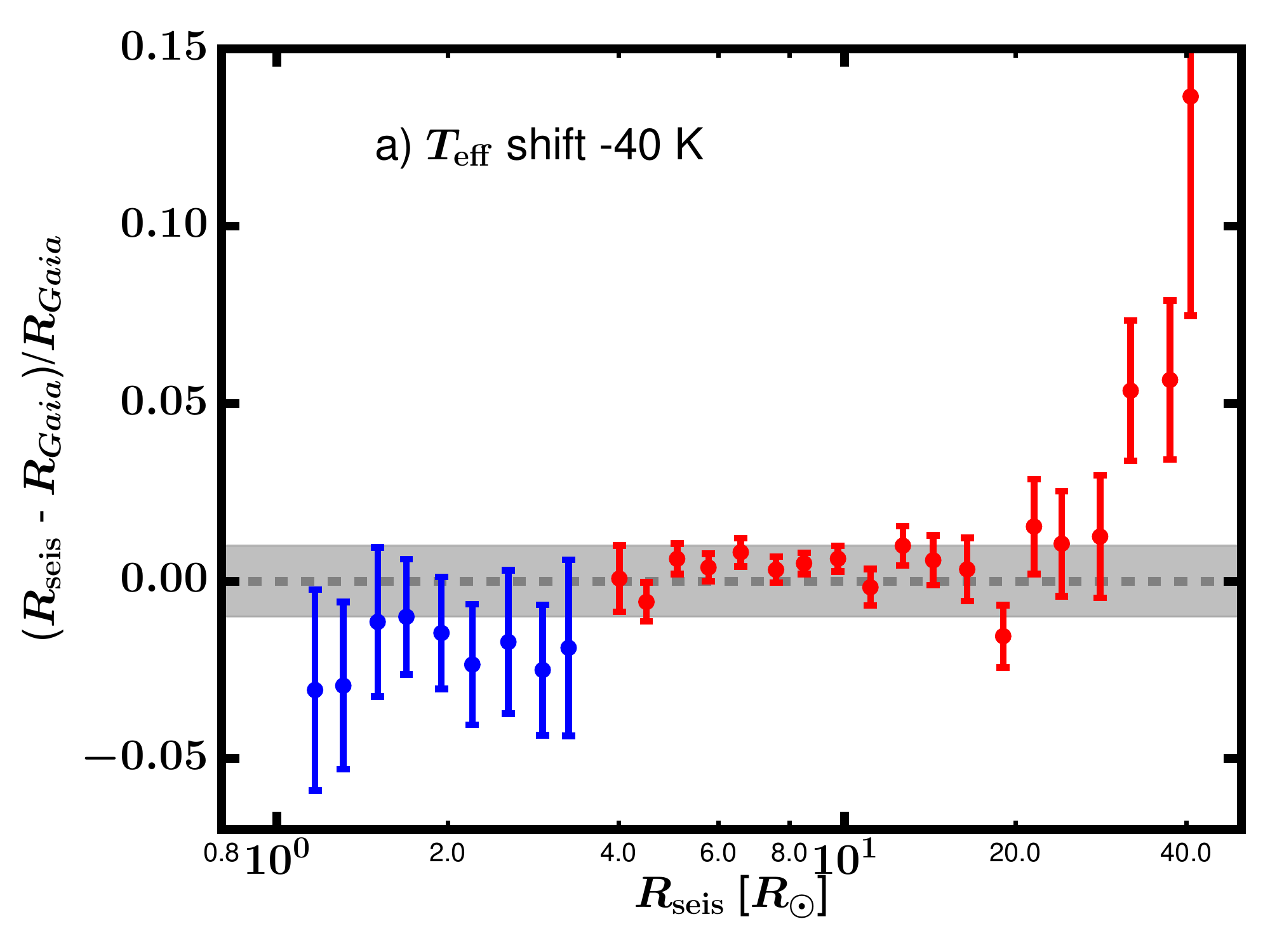}\\
\includegraphics[width=0.49\textwidth]{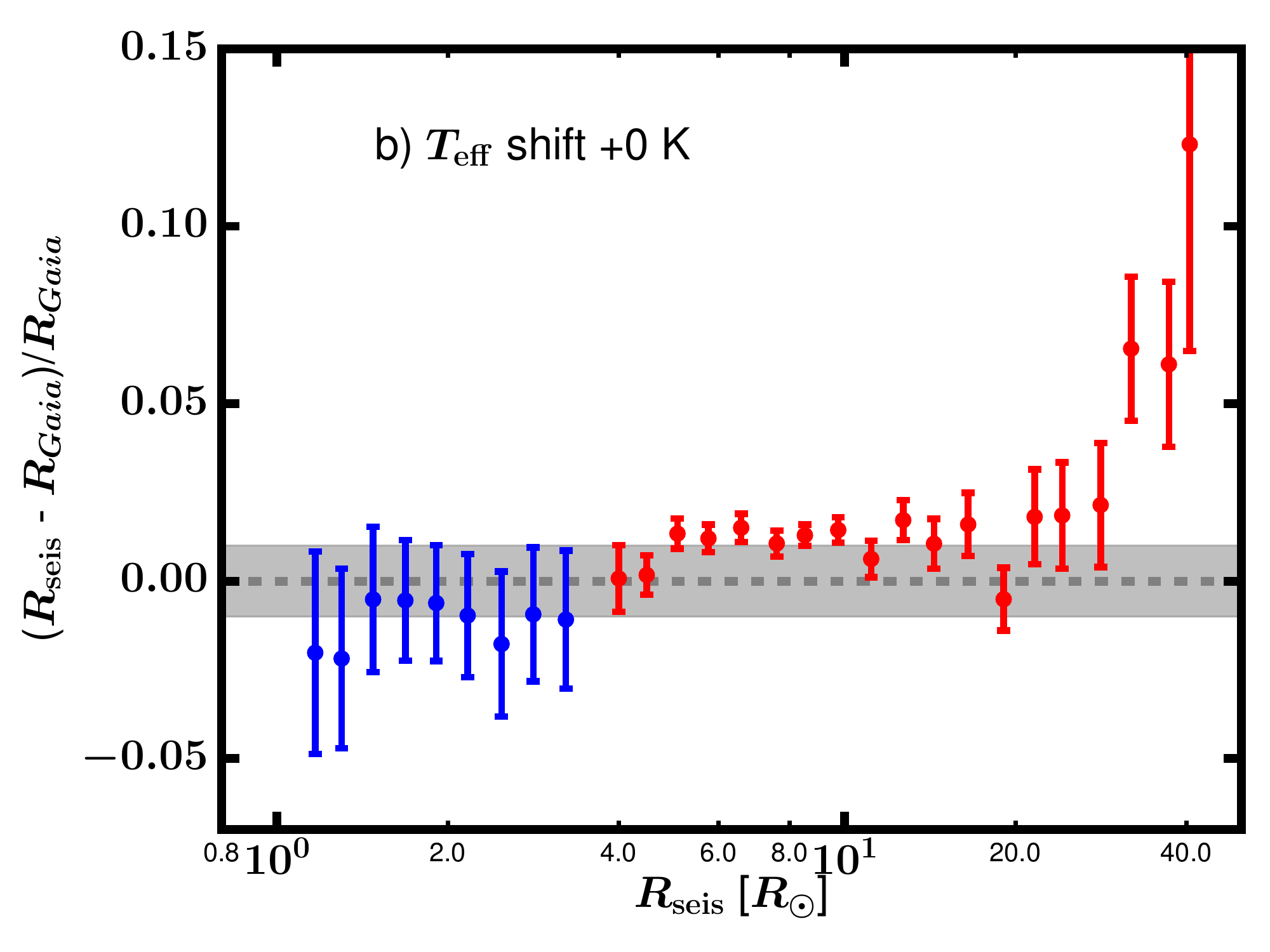}\\
\includegraphics[width=0.49\textwidth]{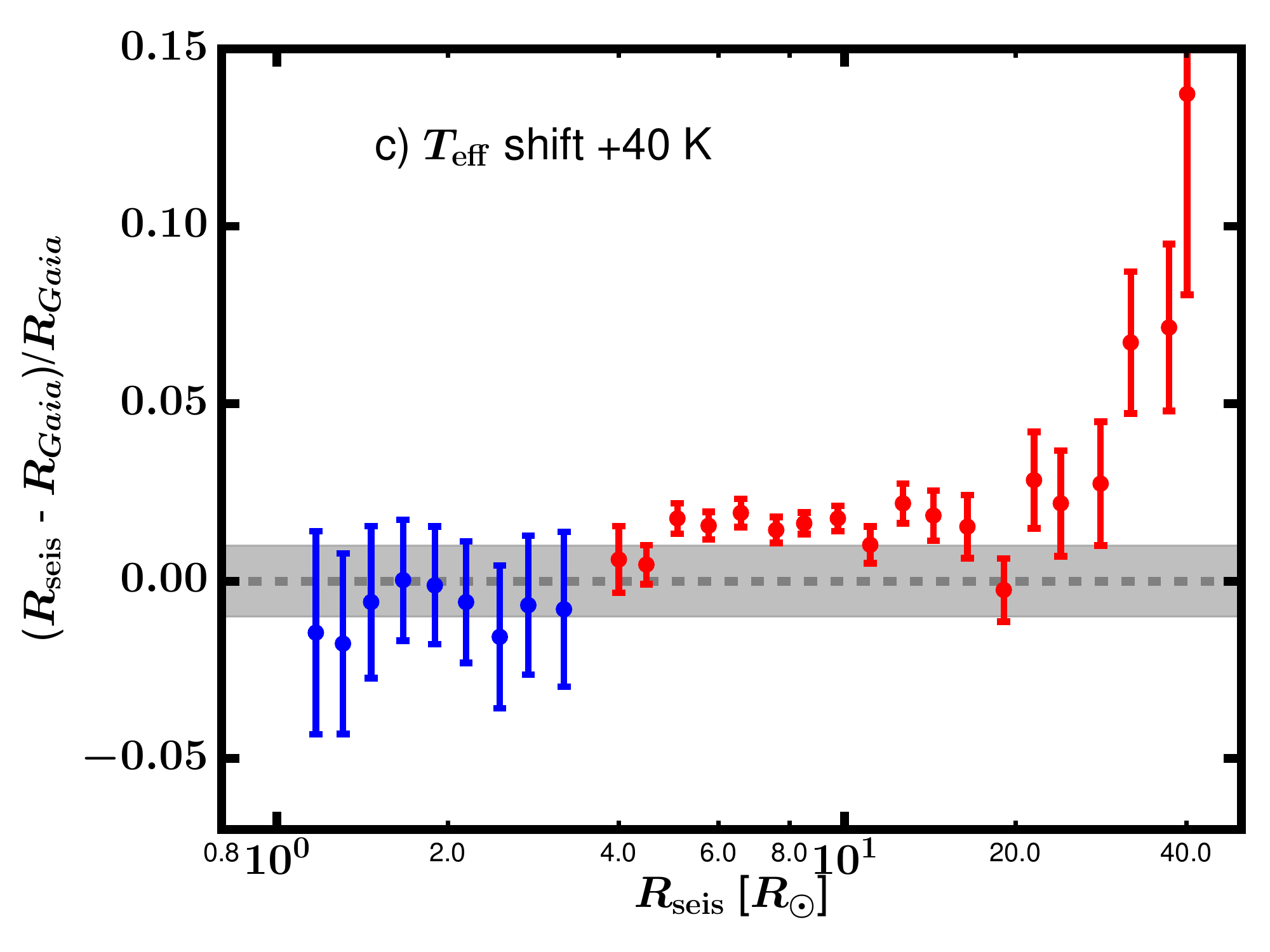}\\
\caption{A close-up of Figure~\ref{fig:radb}c, but also including
  dwarfs/subgiants. The red error bars are binned medians and the
  errors on the binned medians for the giant (red) and dwarf/subgiant
  (blue) samples. The grey band indicates the $\pm 1\%$ agreement
  region. The agreement between asteroseismic and {\rm Gaia} radii is good to within $2\% \pm 2\% {\rm\, (syst.)}$ for dwarfs, subgiants, and giants. Panel a shows the radius agreement if the APOGEE temperature scale is shifted downward by a $2\sigma$ systematic uncertainty on the temperature scale of 40K, panel b shows the radius agreement with the APOGEE temperature scale unchanged, and panel c shows the radius agreement with the APOGEE temperature scale shifted upward by 40K.}
\label{fig:money}
\end{figure}

First and foremost, there is a hint of a differential trend in the radius agreement between $0.8\rsun \lesssim R \lesssim 30\rsun$, which can be seen in Figure~\ref{fig:money}b. Although adjacent radius regimes yield radius correction factors that are statistically consistent with each other (e.g., the flat trend among just giants with $R < 30\rsun$ seen in Figure~\ref{fig:money}b), when considering the radius correction factor required for dwarfs/subgiants ($a_1 = 0.979 \pm 0.005$) and for stars with $10 \rsun < R <   30\rsun$ ($a_3 = 1.019 \pm 0.0060 {\rm\, (rand.)} \pm 0.020 {\rm\, (syst.)}$), they are not statistically consistent with each other at the $5\sigma$ level. One explanation of this trend with radius would be a variation of the underlying physics determining the relationship between asteroseismic frequencies and stellar parameters as a function of radius. Such trends are supposed to be removed by $f_{\dnu}$, but small inadequacies in $f_{\dnu}$ could result in radius-dependent asteroseismic radius errors. This differential trend could also be caused by small systematic trends in the underlying measurements. For instance, small radius-dependent $\numax$ trends are noted by \cite{pinsonneault+2018}; it is also feasible that there exists a small temperature offset between APOGEE dwarf and giant temperature scales. The second trend of note is that the asteroseismic radius scale appears to increasingly over-predict radii compared to {\it Gaia} for $ R \gtrsim 30\rsun$. The statistical significance of this trend is convincing in the sense that there is a bona fide radius inflation, but further work must be done to understand the upper giant branch asteroseismic radius scale --- both observationally and theoretically --- before commenting further on it. These trends are statistically significant, even when perturbing the temperature scale, as we note in \S\ref{sec:choices}.

\subsection{Recommended asteroseismic radius scale}
According to our model for asteroseismic radius correction factors, dwarfs and subgiants have an asteroseismic radius scale that is too small at the $2\%$ level, compared to the {\it Gaia} radius scale. As we noted in \S\ref{sec:abs_results}, the effect is not statistically significant, because it falls within the combined random and systematic uncertainty budget. The effect is reversed among giants, in the sense that stars both below and above the red clump radius ($R \sim 10\rsun$) indicate an {\it inflation} of the asteroseismic radius scale above the {\it Gaia} radius scale at the $2\%$ level. We can interpret these radius scale disagreements as consistent with errors in some combination of bolometric correction, extinction, temperature, the APOKASC-2
asteroseismic radius calibration, or the {\it Gaia} zero-point, which in total allow for systematic shifts in the radius agreement at the $2\%$ level.
We therefore do not recommend specific corrections to the asteroseismic red giant radius scale, but rather conclude
that the giant asteroseismic radius scale, like that of
dwarfs/subgiants, is consistent with the {\it Gaia} radius scale to
within $2\% \pm 2\%\mathrm{\,(syst.)}$. The most evolved giants have asteroseismic radii
that are inflated still further --- by $9\% \pm 2\% \mathrm{\,(syst.)}$.

\begin{figure}
\includegraphics[width=0.5\textwidth]{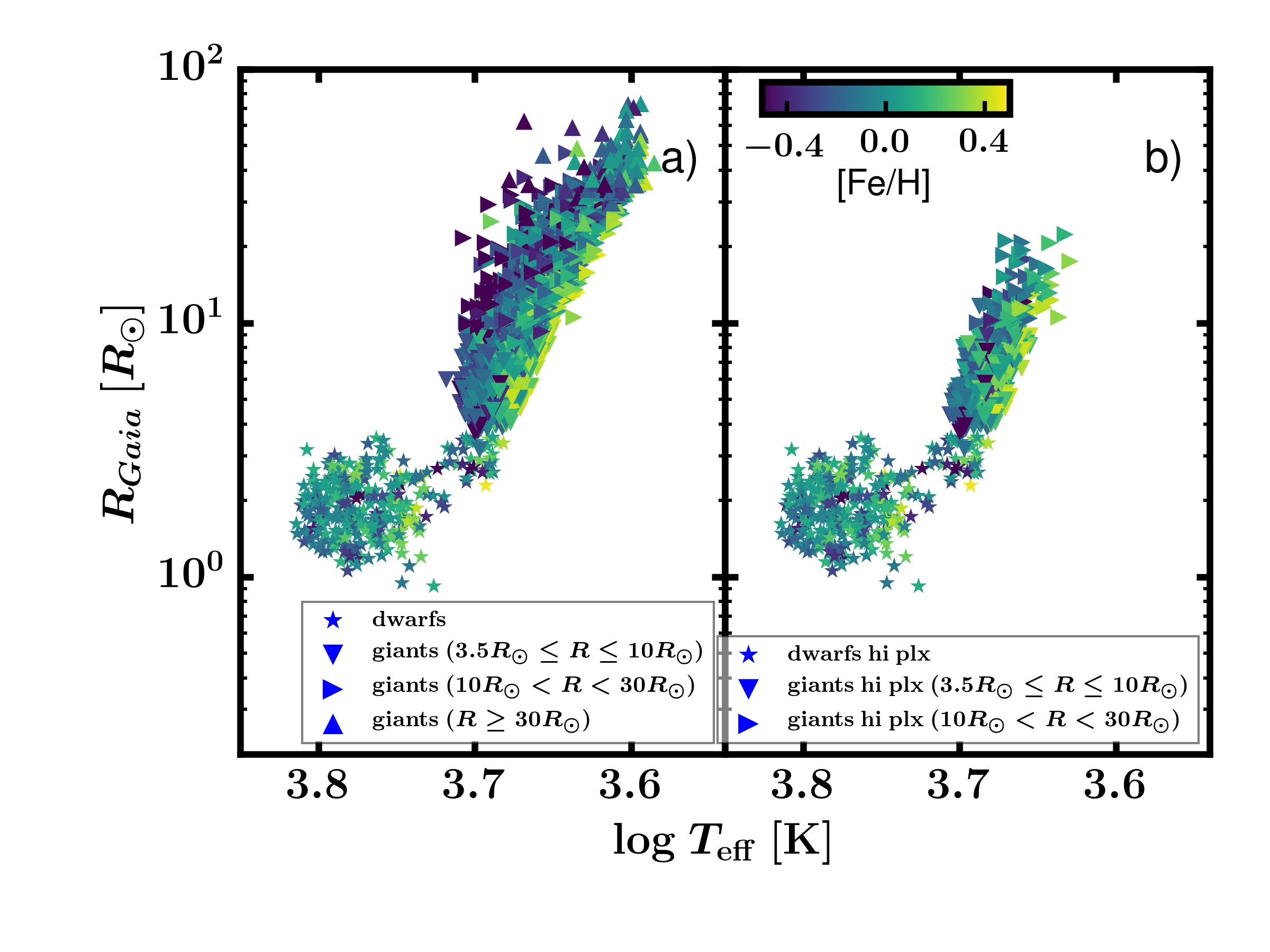}
\caption{{\rm Gaia} radii as a function of temperature for the full giant \& dwarf/subgiant samples (left) and the high-parallax sub-sample (right) used in this work, divided into the four different asteroseismic radius regimes we consider. These radii are excerpted in Table~\ref{tab:radii} in the column $R_{\ks{\rm,\ MIST}}$.}
\label{fig:rad}
\end{figure}

Table~\ref{tab:radii} contains the {\it Gaia} radii we have derived in this work. We provide both radii
corrected for the {\it Gaia} parallax zero-point, and radii that have not been
corrected. Note that a systematic uncertainty of $1.8\%$ should be adopted for the corrected radii, which is smaller than our $2\%$ systematic uncertainty on the ratio of {\it Gaia} and asteroseismic radii because of the smaller temperature dependence of the {\it Gaia} radii compared to the ratio of the two radius scales. The uncorrected {\it Gaia} radii are provided to use in conjunction with a custom {\it Gaia} zero-point, and whose systematic uncertainty would be $1.6\%$, without taking into account systematics due to not correcting for the {\it Gaia} parallax zero-point. The parallax zero-point--corrected radii are plotted in Figure~\ref{fig:rad} as a function of temperature for both the full sample (panel a) and the high-parallax sub-sample (panel b).

\begin{table*}
    \begin{tabular*}{\textwidth}{cccccc}
KIC & $R_{\ks{\rm,\ MIST}} [\rsun]$ & $\sigma_{R_{\ks{\rm,\ MIST}}} [\rsun]$ &
$R_{\ks{,\rm\ MIST,\ raw}} [\rsun]$ & $\sigma_{R_{\ks{,\rm\ MIST,\ raw}}} [\rsun]$ & flags
\\ \hline
 11400880 &   9.75 &   0.71 &  11.08 &   0.88 & 20 \\ \hline
  6587865 &  21.63 &   1.50 &  25.64 &   2.03 & 30 \\ \hline
  5007332 &   6.79 &   0.44 &   7.40 &   0.50 & 20 \\ \hline
  5039087 &  21.98 &   2.39 &  31.30 &   4.61 & 30 \\ \hline
  4832196 &  16.61 &   1.24 &  19.78 &   1.71 & 30 \\ \hline
 10220213 &   4.38 &   0.23 &   4.54 &   0.24 & 21 \\ \hline
 10669876 &  13.12 &   0.62 &  14.14 &   0.69 & 30 \\ \hline
  4139784 &  10.04 &   0.43 &  10.72 &   0.47 & 30 \\ \hline
  3443483 &   6.33 &   0.28 &   6.65 &   0.30 & 20 \\ \hline
  6383574 &  23.42 &   1.38 &  27.17 &   1.75 & 30 \\ \hline
    \end{tabular*}
\caption{A subset of our recommended {\rm Gaia} radii, $ R_{\ks{\rm,\ MIST}}$, and
  their $1\sigma$ random errors, the full list of which is available online. We also include {\rm Gaia}
  radii that have been computed without correcting the {\rm Gaia}
  parallaxes, $R_{\ks{,\rm\ MIST,\ raw}}$. The listed uncertainties do not include systematic contributions to the uncertainties: there is a $1.8\%$ systematic uncertainty on the zero-point--corrected {\rm Gaia} radii and a $1.6\%$ systematic uncertainty on the uncorrected {\rm Gaia} radii, which does not account for the error induced by not correcting for the {\it Gaia} parallax zero-point. Flags are two digits in length: the first digit indicates to which of the four asteroseismic radius bins the star belongs (either 1, 2, 3, or 4 corresponding to Equation~\ref{eq:broken}); and the second digit is 1 if the star is a part of the high-parallax sub-sample, or 0 otherwise.}
\label{tab:radii}
\end{table*}

\subsubsection{Scaling relations as a function of metallicity for {\rm [Fe/H]} $\geq -1$}
\label{sec:met}
\begin{figure}
\includegraphics[width=0.5\textwidth]{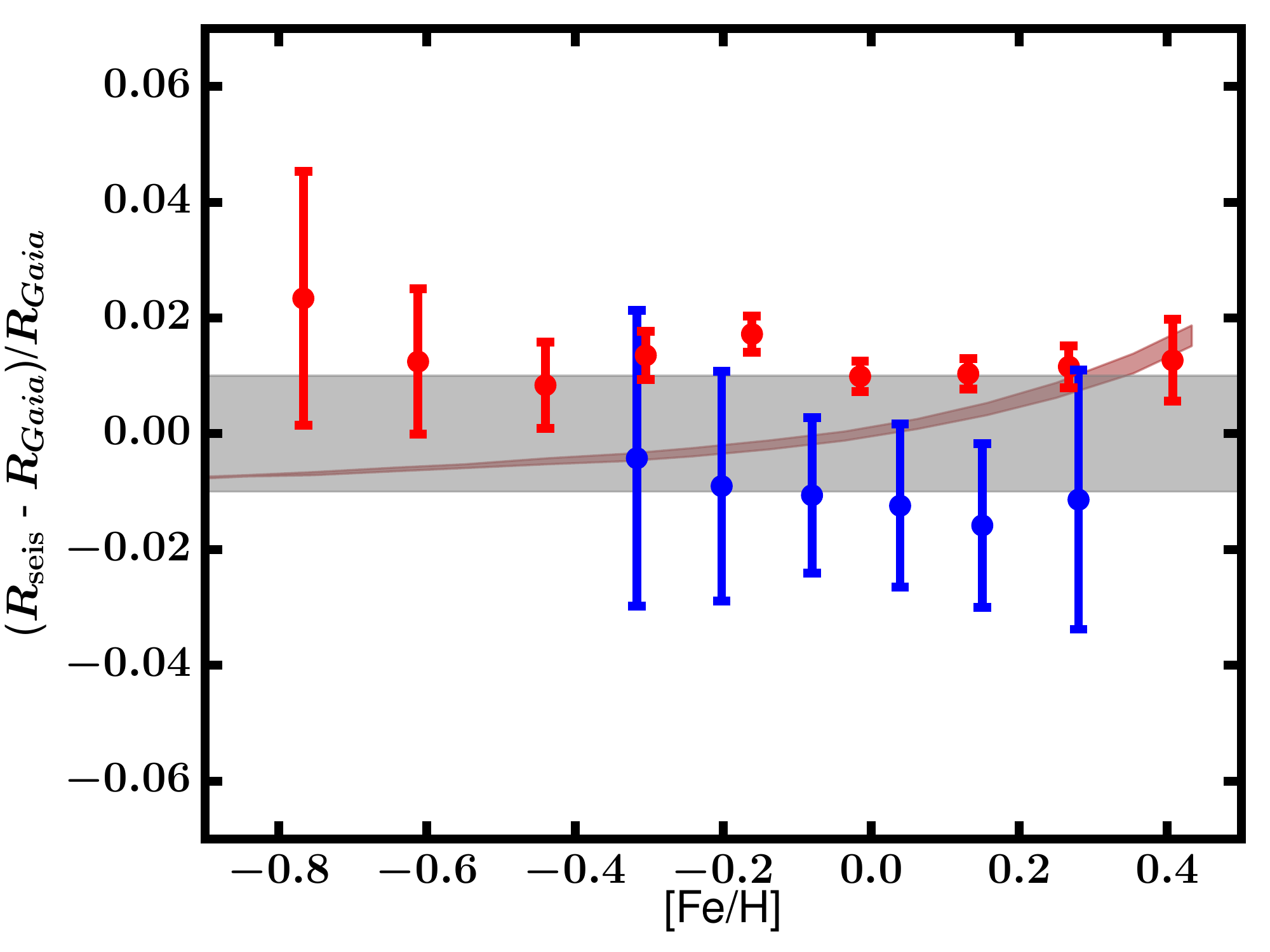}
\caption{Difference between asteroseismic and {\rm Gaia} parallax as a
function of metallicity, after correction using our adopted {\rm Gaia}
parallax zero-point, but with no asteroseismic radius correction factors applied. The median and error on the median radius agreement in bins of metallicity for giants are shown as red error bars and for dwarfs/subgiant as blue error bars. The grey band indicates an agreement between the radius scales to within $\pm 1\%$. The brown band indicates the expected disagreement from \cite{viani+2017a} between the red giant radius scales with and without taking into account a molecular weight term. See \S\ref{sec:met} for details.} 
\label{fig:met}
\end{figure}

Based on the argument that scaling relations depend on the sound
speed, and that the sound speed depends on molecular
weight, \cite{viani+2017a} have proposed that the $\numax$ asteroseismic scaling
relation (Equation~\ref{eq:dnuscaling}) should depend on metallicity. This theory would predict that $f_{\numax}$ in Equations~\ref{eq:scaling3} \&~\ref{eq:dnuscaling} would be non-unity and a function of metallicity. We can test this prediction
with our data, by showing the parallax difference as a function of
metallicity, as we do in Figure~\ref{fig:met}. Here, we have plotted the observed radius agreement as a function of [Fe/H], and have included the expected error in asteroseismic radius for the giants in the sample due to not including a molecular weight term in the scaling relations, according to Equation 21 of \cite{viani+2017a} (brown band).  The width of this band is due to the spread in [$\alpha$/Fe], which we take from the APOKASC-2 catalogue. We compute the molecular weight according to $\mu = 4/(3X + 1)$, assuming a helium enrichment of $\Delta Y/\Delta Z = 1$, a primordial Helium abundance of $Y=0.248$, $Z_{\odot} = 0.02$, and for each star in the sample, $Z = 10^{ 0.977 \mathrm{[M/H]} - 1.699}$ \citep{bertelli+1994a}, where $\mathrm{[M/H]} = \mathrm{[Fe/H]} + \log(0.638 10^{\mathrm{[}\alpha\mathrm{/Fe]}} + 0.362)$ \citep{salaris+1993a}. The primary assumption in this simple implementation of a metallicity-dependent $f_{\numax}$ is that there is a one-to-one relation between metallicity and helium fraction. A spread in intrinsic helium fraction would tend to smear out any trend with metallicity and therefore flatten the predicted effect. In our expression for mean molecular weight, we have also assumed that the gas is neutral in the acoustic radius of the star, which induces an uncertainty in the predicted metallicity-dependent radius error. There should also be an uncertainty due to not considering the adiabatic index in the atmosphere of the star, which will depend on metallicity. Investigating the impact of these effects would require detailed modeling of the stars, which is beyond the scope of this work. With these modeling caveats in mind, across the more than 1 dex spread in metallicity shown in Figure~\ref{fig:met}, we do not see evidence for the predicted metallicity effect. Indeed, the data are consistent with having no trend with metallicity to within $0.5\%$ per dex for giants and $1.1\%$ per dex for dwarfs/subgiants, based on least-squares fitting. Taking into account the $2\%$ systematic uncertainty in our radius comparison does not change this conclusion, because the systematic is insensitive to metallicity, and therefore would tend to shift all of the data shown in Figure~\ref{fig:met} up or down. Until such a time as the intrinsic scatter in helium enrichment can be determined, which, at this point, hinders a comparison between the theoretical metallicity trend and the observed radius agreement, we conclude that the asteroseismic scaling relation radius does not require a metallicity term to within the precision afforded to us by our data set.

\subsubsection{Scaling relations for {\rm [Fe/H]} $< -1$}
\begin{figure}
\includegraphics[width=0.5\textwidth]{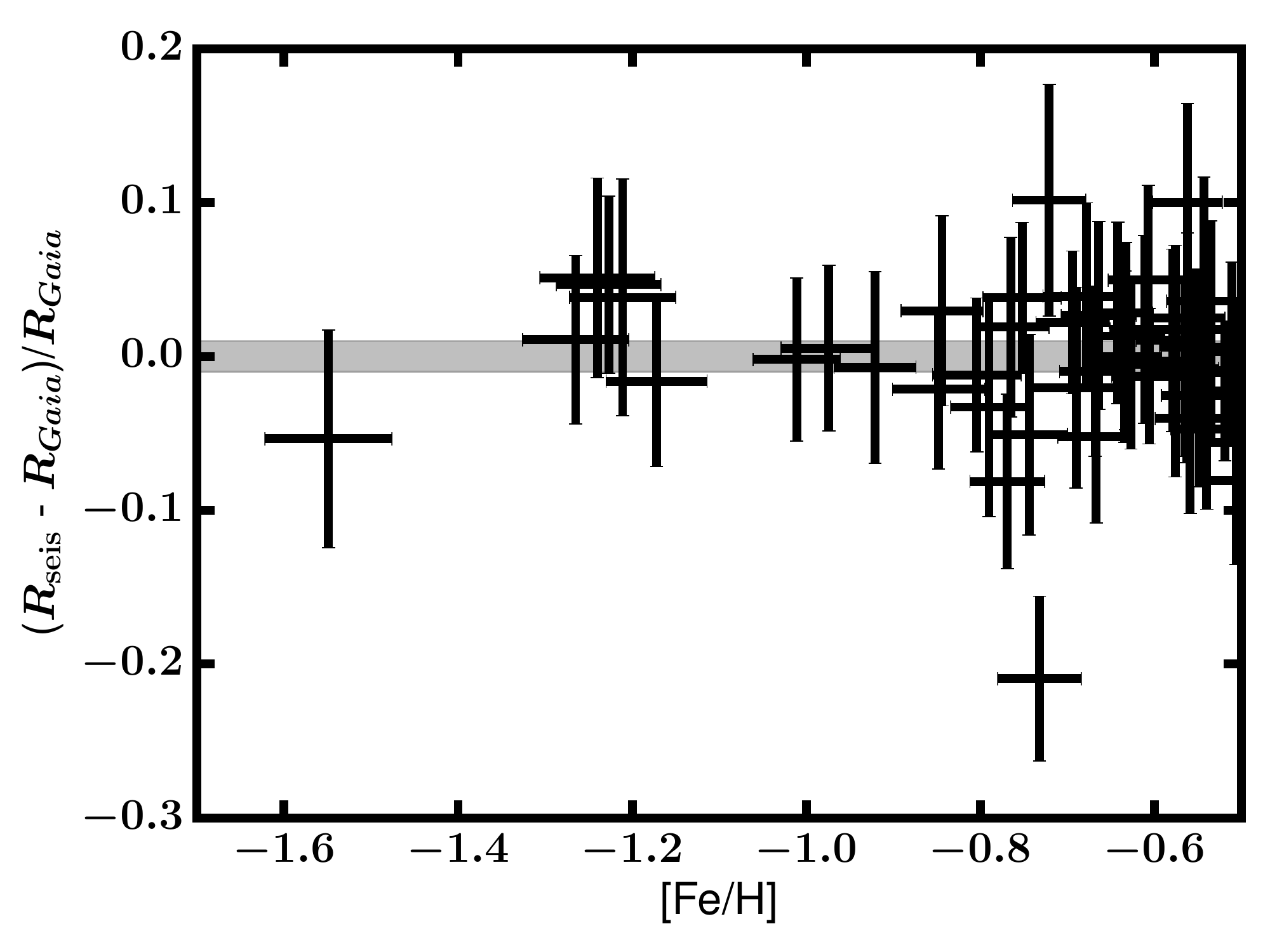}
\caption{Fractional difference between asteroseismic and {\rm Gaia} radius as a
function of metallicity for low-metallicity stars with $\rscal \leq
10 \rsun$. A grey band corresponding to $\pm 0.01$ has been added to
guide the eye. There is no statistically significant evidence for a metallicity-dependent asteroseismic radius error for [Fe/H] $ < -1.0$. See \S\ref{sec:met} for details.}
\label{fig:lo_met}
\end{figure}

Motivated by the observation in \cite{epstein+2014} that halo stars have asteroseismic masses that appear to be inflated compared to the masses expected from stellar models, we discuss here the asteroseismic radius and mass scale in the halo metallicity regime ([Fe/H] $< -1$). There seems to be no significant disagreement in radius space for the most metal-poor stars, which we show in
Figure~\ref{fig:lo_met}. Here, we have only shown the stars below the
red clump ($\rscal \leq 10\rsun$) as black error bars, to disambiguate metallicity-dependent effects and radius scaling relation effects that we find in the most evolved stars (see \S\ref{sec:diff2}). To isolate the metallicity effect, the $a_2$ radius correction factor is applied. When correcting for
the radius correction factor derived from the high-parallax sub-sample at all metallicities as well as the parallax offset using the {\it Gaia} zero-point model from \cite{zinn+2019}, which includes a color term,
the radius anomaly of the eight 
stars with [Fe/H] $< -1.0$ and $\rscal \leq 10\rsun$ is $1.02 \pm 0.02 ({\rm rand.}) \pm 0.02 ({\rm syst.})$ and does thus not deviate from unity. The color term ($d$ in Equation~\ref{eq:zp}), however,
will tend to correct for metallicity effects, as well, if
present. Even when only correcting the {\it Gaia} parallaxes using the radius correction factor and a global
offset term, $c$, the anomaly is still not statistically
significant, at $1.02 \pm 0.02 ({\rm rand.}) \pm 0.02 ({\rm syst.})$. For this reason, there does not appear to be a problem with the asteroseismic radius scale at low metallicity.

We can also infer the corresponding inflation in mass space, by combining the mass scaling relation, $\frac{M_{\mathrm{seis}}}{\msun} \approx \left( \frac{\numax}{f_{\numax} \numaxsun}\right)^{3}  \left( \frac{\dnu}{f_{\dnu} \dnusun}\right)^{-4} \left(\frac{\teff}{\teffsun}\right)^{3/2}$, with a {\it Gaia} radius to yield a {\it Gaia} mass, which depends on both parallax and $\dnu$: $\frac{M_{Gaia}}{\msun} \approx \left( \frac{\dnu}{f_{\dnu} \dnusun}\right)^{2} \left(\frac{R_{Gaia}}{\rsun}\right)^{3}$. The assumption here is that $f_{\dnu}$ corrects the scaling relation completely so that $M_{Gaia}$ is unbiased, whereas the asteroseismic mass has an additional dependence on $\numax$; looking at the ratio of {\it Gaia} to asteroseismic radius for a low-metallicity sample would reveal a metallicity-dependent $f_{\numax}$. We have already inferred in \S\ref{sec:results} that there is a statistically insignificant but non-zero asteroseismic radius correction factor for stars with $R \leq 10\rsun$ of $a_2 = 1.015$ averaged over the entire sample (with relatively high metallicities, mostly $-0.2 < \mathrm{[Fe/H]} < 0.2$). We find for these eight stars $\langle M_{Gaia} / M_{\mathrm{seis}} \rangle = 0.94 \pm 0.08 {\rm \,(rand.)} \pm 0.07 {\rm \, (syst.)}$ when correcting only for the radius correction factor and the {\it Gaia} global zero-point, and $\langle M_{Gaia} / M_{\mathrm{seis}} \rangle = 0.96 \pm 0.08 {\rm \,(rand.)} \pm 0.07 {\rm \, (syst.)}$ when also accounting for the color and magnitude terms. These ratios depart mildly from unity, but not strongly. Here, we have corrected the $\numax$ scale for the radius inflation effect we note in this paper, which lowers the asteroseismic mass scale by $4.5\%$ given $R_{\mathrm{seis}} \propto \numax$ and $M_{\mathrm{seis}} \propto \numax^{3}$. The mass ratio we find is in agreement with that from \cite{epstein+2014}, who found a mass ratio of $0.89 \pm 0.04$ when comparing halo and thick disk masses expected from stellar models to asteroseismic masses corrected with $f_{\dnu}$ according to the \cite{white+2011} prescription. The strong temperature dependence, $M_{Gaia} / M_{\mathrm{seis}}  \propto T^{-15/2}$, means that the ratio is particularly sensitive to temperature scale systematics, and so improvement upon these estimates of a metallicity effect may prove difficult even using a larger sample of halo stars.

\section{Discussion}
\label{sec:discussion}

\subsection{Comparison with literature}
\begin{figure}
\includegraphics[width=0.5\textwidth]{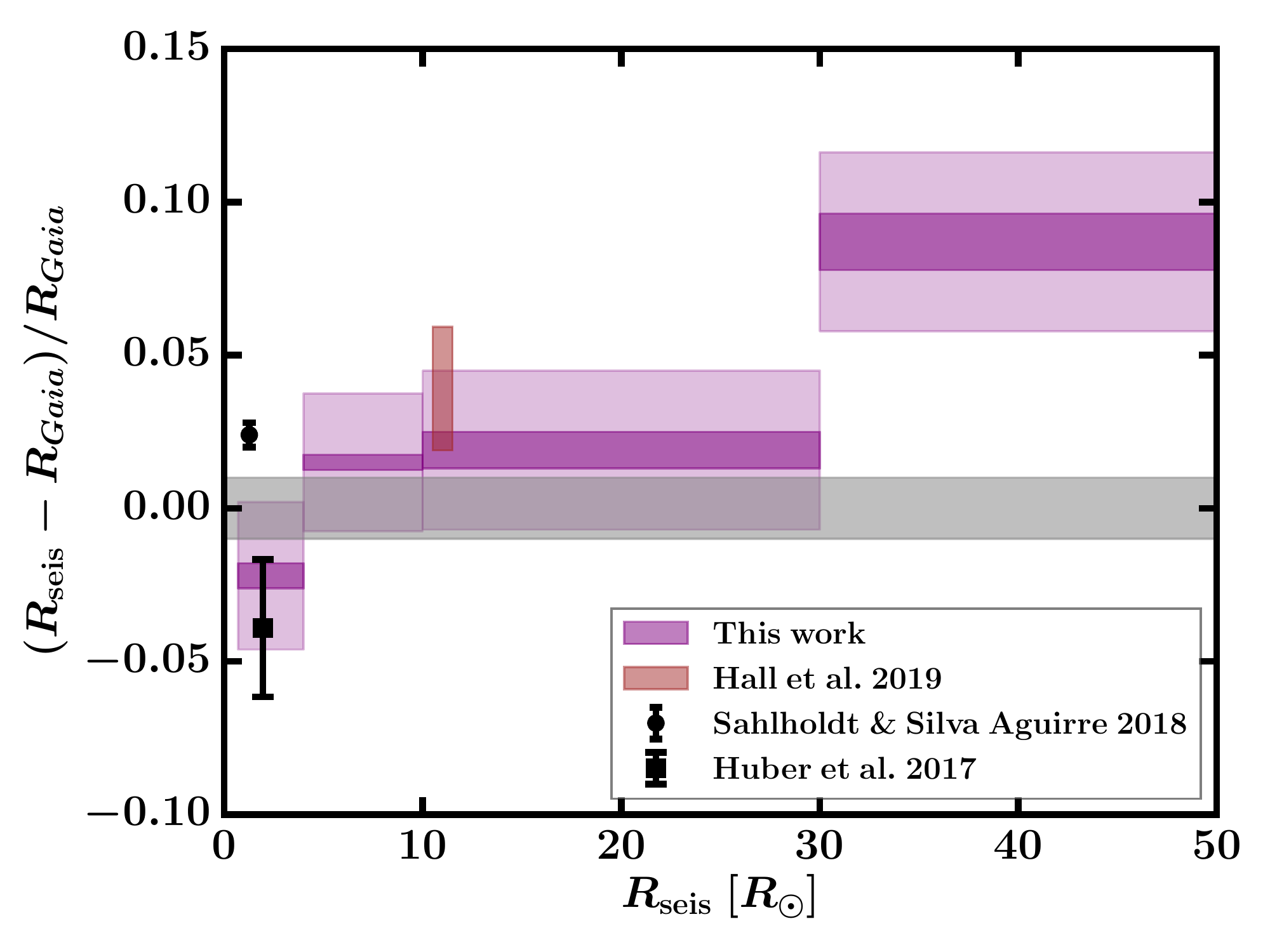}
\caption{Comparison of asteroseismic-{\rm Gaia} radius agreement among
  literature estimates and this work. The dark purple bands indicate
  the best-fitting radius correction factors that would bring
  asteroseismic radii into agreement with {\rm Gaia} radii
  (Table~\protect{\ref{tab:results}}), and the light purple bands
  indicate the $1\sigma$ systematic possible due to uncertainties in
  the luminosity scale, the temperature scale, and the asteroseismic
  radius scale. A gray band corresponding to $\pm 0.01$ has been added to
  guide the eye. See \S\ref{sec:marquee} for details.}
\label{fig:marquee}
\end{figure}
\subsubsection{Constraints from {\it Gaia}}
\label{sec:marquee}
We compare in Figure~\ref{fig:marquee} the radius agreement we find in this work to recent work comparing the {\it Gaia} radius scale to the asteroseismic radius scale. First we consider the result from \cite{hall+2019}, who performed a hierarchical Bayesian analysis of the red clump absolute magnitude in the $\ks$- and {\it Gaia} $G$-bands using both an asteroseismic luminosity and a {\it Gaia} luminosity. Using their best-fitting {\it Gaia} absolute luminosity in the $\ks$-band of $    \mu_{\mathrm{RC,\ Gaia}} = -1.634 \pm 0.018$ (which uses an uninformative prior on the {\it Gaia} parallax zero-point) and their best-fitting value using asteroseismology and APOKASC-2 temperatures of $    \mu_{\mathrm{RC,\mathrm{seis}}} = -1.693 \pm 0.003$, yields a radius agreement that is statistically consistent with the one inferred by us for RGB stars near the radius of the clump $R \sim 10\rsun$. The absolute magnitude constraint from \cite{hall+2019} is not a pure radius constraint, however, as the absolute magnitude depends on the luminosity and thus the temperature of the star. On the asteroseismic side, \cite{hall+2019} uses temperatures either from APOKASC-2 or from \cite{mathur+2017}. The former is the same temperature scale we adopt in this work, and so the red clump asteroseismic-{\it Gaia} absolute magnitude agreement from \cite{hall+2019} using the APOKASC-2 red clump stars would be an appropriate point of comparison to our constraints on the radius agreement along the first-ascent giant branch. However, the {\it Gaia} red clump absolute magnitude estimate from \cite{hall+2019} is based on a sample of stars from the asteroseismic analysis of \cite{yu+2018}, which have temperatures from \cite{mathur+2017}, which are hotter on average than those from APOKASC-2. Taking into account this temperature effect results in a range of possible radius agreement on the red clump, which is shown in Figure~\ref{fig:marquee} (the \citealt{hall+2019} result has been placed at a representative location on the abscissa in Figure~\ref{fig:marquee} of $R=11\rsun$ and with an spread of $1\rsun$, according to their Figure 2).  We see agreement within the uncertainty between the \cite{hall+2019} radius comparison and the result from this work. \cite{hall+2019} postulates that the difference they find between asteroseismic and {\it Gaia} absolute magnitudes could be explained by a systematic offset of $-70$K in the spectroscopic temperature scale. Systematic differences among uncalibrated spectroscopic temperature scales can indeed disagree at this level. However, as we note in \S\ref{sec:teff_syst} the APOGEE temperature scale has a $1\sigma$ systematic uncertainty of $20$K because it has been calibrated to the IRFM temperature scale. \cite{hall+2019} also finds that the $f_{\dnu}$ choice for red clump stars can significantly shift the red clump absolute magnitude scale. In this sense, a percent level offset between the asteroseismic radius scale of red giants and red clump stars is easily accommodated by the systematics in red clump models used to compute $f_{\dnu}$ \citep[e.g.,][]{pinsonneault+2018,an+2019,hall+2019}.

\begin{figure}
\begin{center}
\includegraphics[width=0.5\textwidth]{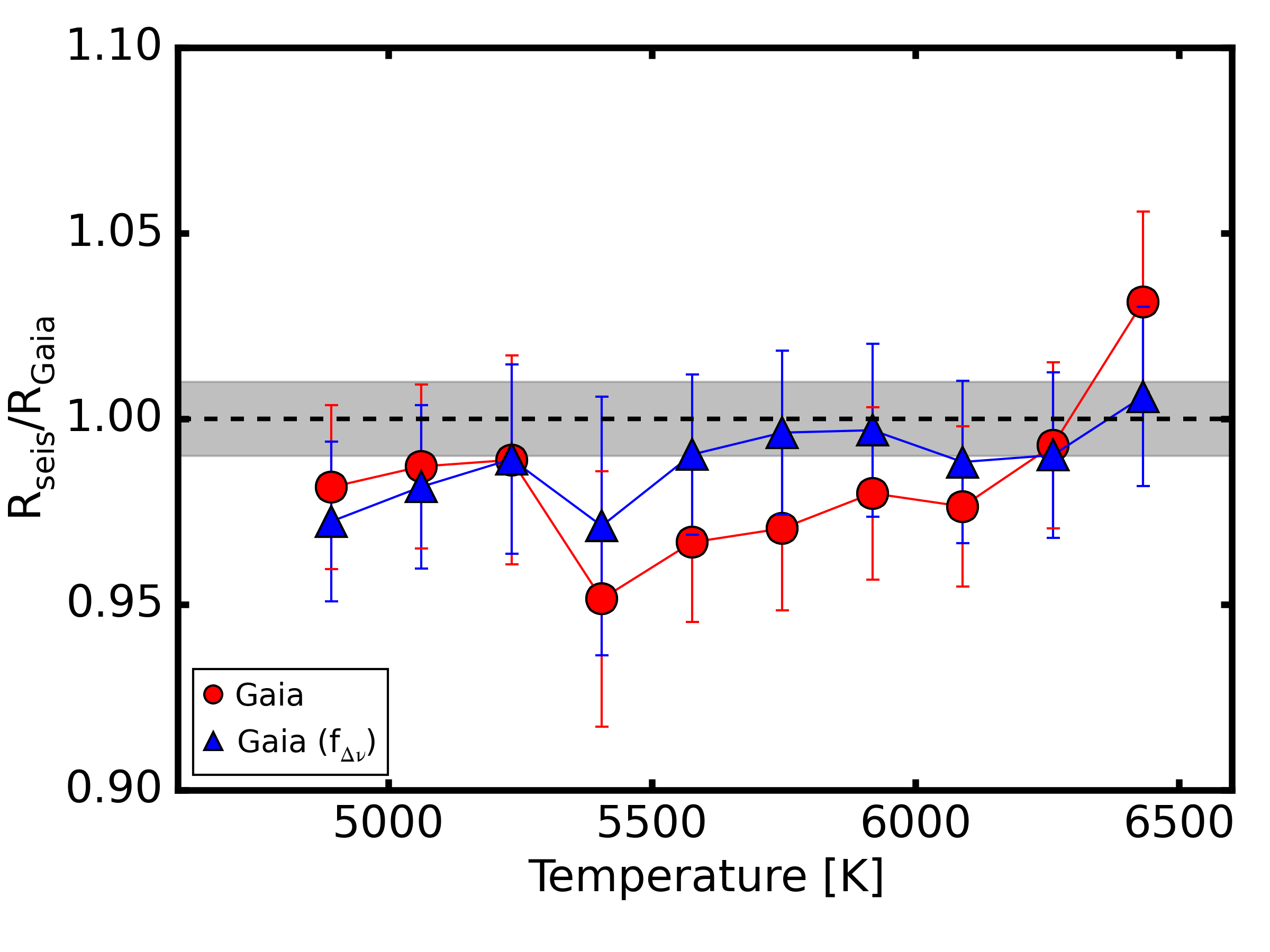}
\caption{Comparison of asteroseismic radii derived from scaling relations to those derived from {\it Gaia} parallaxes, as a function of temperature. Red circles and blue upward triangles show our dwarf/subgiant sample without and with the use of $f_{\dnu}$. Error bars indicate scatter in the median. The grey band indicates agreement to within $1\%$.}
\label{fig3}
\end{center}
\end{figure}

\cite{sahlholdt+2018b} investigated the agreement between asteroseismology and {\it Gaia} radius scales among dwarfs and subgiants using {\it Gaia} DR2 parallaxes. Using scaling relations corrected according to \cite{white13}, they found a mean ratio of $\langle R_{\mathrm{seis}}/R_{Gaia} \rangle = 1.024 \pm 0.004$ (plotted in Figure~\ref{fig:marquee}). An additional set of asteroseismic scaling relation radii were computed using an additional set of surface corrections following \cite{ball_gizon_2014}, and which yielded a mean $\langle R_{\mathrm{seis}}/R_{Gaia} \rangle = 1.002 \pm 0.004$. Both of these estimates are mildly discrepant with our estimates and those of \cite{huber+2017} in the dwarf and subgiant regime. This could be due to the simple polynomial expansion in temperature that \cite{white13} employs to parametrize $f_{\dnu}$ as opposed to the grid-based interpolation scheme from BeSPP. The asteroseismic data from \cite{sahlholdt+2018b} are also not calibrated to be on the cluster mass scale (as are the data we use in this work), which could help to explain the tension.  \cite{sahlholdt+2018b}  also found deviations
of $\pm 3\%$ at the extreme ends of their sample's temperature distribution, near
$5400K$ and $6600K$ (their Figure 4c). When we view our dwarf radius comparison as a function
of temperature, shown in Figure~\ref{fig3}, we see a similar effect at $\sim 5400K$, but not at hotter temperatures. We believe that the lack of any
trends beyond the $1\%$ level with temperature at hotter temperatures is a result of a difference in our adopted $f_{\dnu}$.

Finally, Figure~\ref{fig:marquee} also shows the mean and error on the mean of the radius agreement from \cite{huber+2017}, who worked with {\it Gaia} DR1 and the same dwarf/subgiant asteroseismic sample used in this work. These results are consistent with ours, though with a larger uncertainty due to the less precise parallaxes in {\it Gaia} DR1.

\begin{figure}
\begin{center}
\includegraphics[width=0.5\textwidth]{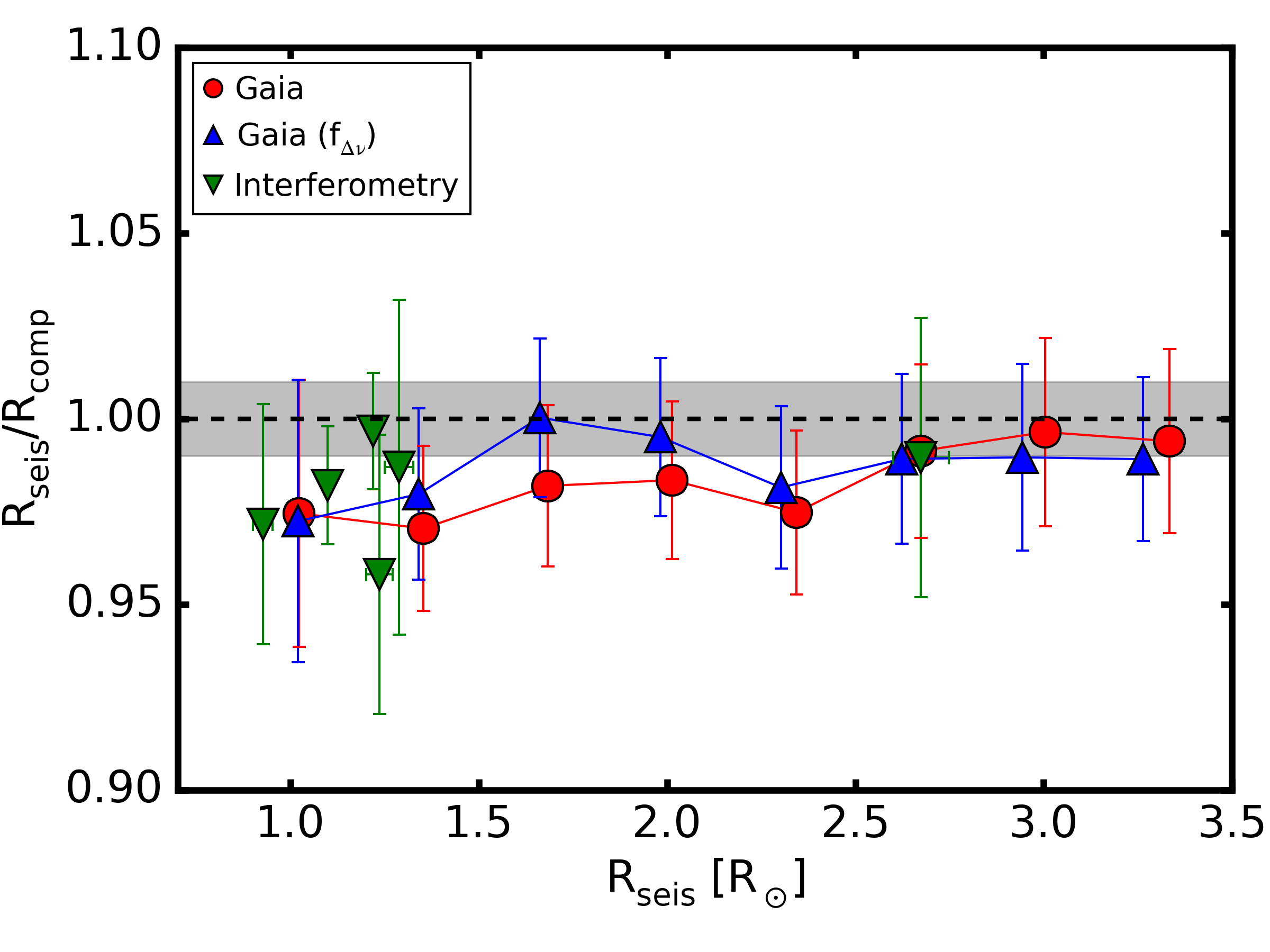}
\caption{Comparison of asteroseismic radii derived from scaling relations with radii derived from three methods in the dwarf/subgiant radius regime ($R < 3.5\rsun$). Red circles and blue upward triangles show our dwarf/subgiant sample without and with $f_{\dnu}$. We also show stars with interferometrically measured radii  \citep[green triangles,][]{huber12,white13,johnson14}. Error bars indicate scatter in the median. The grey band indicates agreement to within $1\%$.}
\label{fig2}
\end{center}
\end{figure}

To analyze our dwarf/subgiant radius comparison in more detail, we reproduce Figure 10 of \citet{huber+2017} in Figure \ref{fig2} by comparing the {\it Gaia} results to independent comparisons from interferometry \citep[e.g.][]{huber12,white13}. The $\approx$\,5\% offset for subgiants identified by \citet{huber+2017} (with asteroseismic radii being smaller) is significantly reduced, suggesting that at least part of that offset may have been caused by an incomplete understanding of the {\it Gaia} parallax systematics in DR1, which would have affected the typically more distant subgiants more than the typically more nearby dwarfs. The largest offsets with {\it Gaia} DR2 are at the $\approx$\,2\,\% level, fully consistent to within 1\,$\sigma$ with the uncertainties for seismic radii derived from scaling relations using corrected $\dnu$ values via $f_{\dnu}$. This excellent agreement strongly suggests that scaling relation radii (using $f_{\dnu}$ according to Equation~\ref{eq:dnuscaling}) are precise and accurate at the $2\% \pm 2\% {\rm\, (syst.)}$ percent level for stars in the range $R \approx 0.8-3.5\,\rsun$. 

Comparing {\it Kepler} first-ascent red giant branch and red clump
asteroseismic parallaxes to {\it Gaia} DR2 parallaxes, \cite{khan+2019}
find agreement between the {\it Gaia} and asteroseismic radius scales within
$\sim 5\%$. We note that our results are not directly comparable
because they do not account for $f_{\Delta \nu}$, and so their level
of agreement between {\it Gaia} and asteroseismic radius scales is an upper
bound. Their results nevertheless confirm our conclusion that the
asteroseismic radius scale is very accurate for red giants.

\subsubsection{Constraints from eclipsing binaries}
The largest study of the red giant asteroseismic radius and mass
scales using eclipsing binaries concluded that the radius scale was
overestimated by $5\%$ compared to the dynamical radius scale
\citep{gaulme+2016}. The latter study examined stars with radii less than
$15\rsun$, and so our results for the smaller-radius stars ($ R \leq 10
\rsun$) are directly comparable. Our
results in this radius regime indicate that the agreement, in fact, is much better than
$5\%$. In that sense, our results accord with indications from
\cite{brogaard+2018} that the temperatures in \cite{gaulme+2016} could
be affected by the blending of the binary systems, therefore biasing the asteroseismic radii. For our sample, however, we use spectroscopic temperatures, which are not sensitive in the same way as photometric estimates are to blending, and we have furthermore selected against binarity using the {\it Gaia} data quality cuts described in \S\ref{sec:gaia}.

\subsection{Dependence on the luminosity and temperature scales}
\label{sec:choices}

In converting asteroseismic radii to parallaxes according to Equation~\ref{eq:plx}, the luminosity scale enters through a dependence on the bolometric flux and distance/parallax, and the temperature enters through the explicit temperature dependence as well as the bolometric correction dependence on temperature. In this section, we discuss in this section checks we have performed to ensure that our adopted luminosity and temperature scales in this work do not bias the radius agreement beyond our systematic uncertainty estimates in \S\ref{sec:bc_syst}.

The observed variations of $a_2$ and $a_3$ using different choices for bolometric correction and extinction are generally within our estimated systematic bolometric correction and extinction error of $1\%$ (\S\ref{sec:abs_results}), when including the random errors quoted on $a_2$ and $a_3$. Interestingly, the agreement between SED and {\it Gaia} radii is closer to unity than the asteroseismic-{\it Gaia} radius comparison. We show in Appendix~\ref{app:fbol} that it is the SED bolometric fluxes that differ the most from the MIST $\ks$-band bolometric corrections among the independent bolometric flux scales we compare to. So whereas the SED bolometric flux scale differs from the one we adopt for our asteroseismic-{\it Gaia} radius comparison by $\sim 4\%$, a difference of $\sim 0.2$ mag in the SED extinctions and those from \cite{rodrigues+2014} that we adopt for our asteroseismic radii compensates to bring the SED radius scale closer to the {\it Gaia} radius scale.

\begin{figure*}
\includegraphics[width=1.0\textwidth]{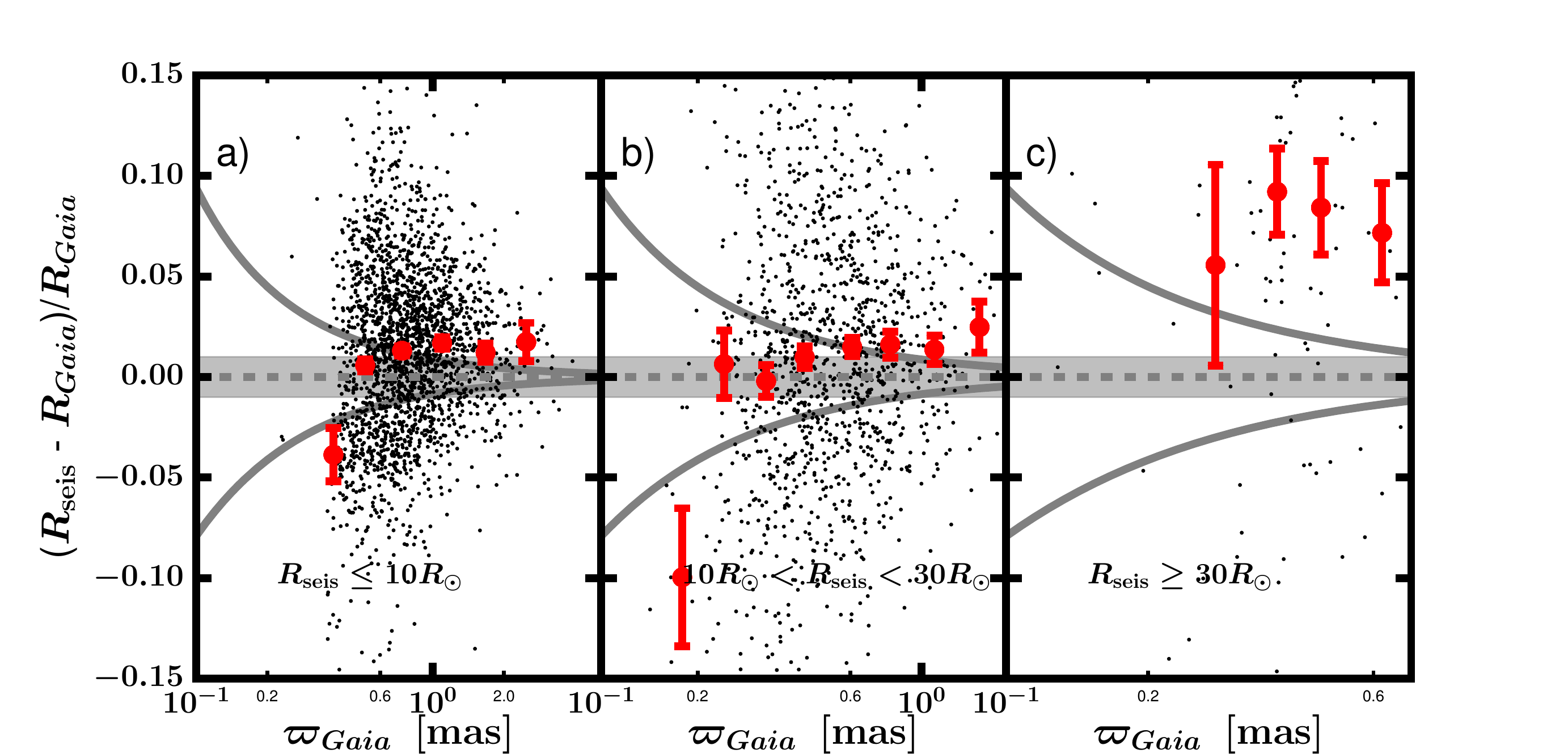}
\caption{The fractional difference between asteroseismic and {\rm
Gaia} scales as a function of {\rm Gaia} parallax for stars with
$\rscal \leq 10\rsun$ (a), $ 10 \rsun < \rscal < 30\rsun$ (b), and
  $\rscal \geq 30\rsun$ (c). A gray band corresponding to $ \pm 0.01$ has
  been added to guide the eye. The solid grey curves show the expected trend with parallax of the fractional radius agreement if our adopted {\rm Gaia} zero-point were shifted by the systematic uncertainty on $c$ of $\pm 8.6\muas$ from \cite{zinn+2019}; the flatness of the grey curves at large parallax indicate large-parallax stars are essentially unaffected by the {\rm Gaia} parallax zero-point correction. We use a high-parallax ($\varpi > 1$mas) giant sub-sample for all but the largest radius regime, $\rscal \geq 30\rsun$, to infer the radius agreement between asteroseismic and {\rm Gaia} scales in this work.}
\label{fig:v_plx}
\end{figure*}

The other component of the luminosity scale involves the parallaxes. The parallax zero-point correction we apply consists of both color- and magnitude-dependent terms ($d$ and $e$ in Equation~\ref{eq:zp}) as well as a global zero-point correction, $c$, with values taken from \citep{zinn+2019}. An argument could be made that the parallax zero-point correction, which is itself constrained by the
asteroseismic data from \cite{zinn+2019}, necessarily enforces agreement between the
asteroseismic and {\it Gaia} radius scales. For reasonable values of the color and magnitude terms in the {\it Gaia} parallax zero-point correction in Equation~\ref{eq:zp}, however, the asteroseismic radii remain consistent
with the {\it Gaia} radii. Figure~\ref{fig:radb}b shows a model without
color and magnitude terms and without radius scale factors $a_2$, $a_3$, and
$a_4$. It is, in this sense, a conservative estimate of the
agreement between asteroseismic and {\it Gaia} radii. This
simplified model is still in excellent agreement with the observed
ratio of asteroseismic to {\it Gaia} radii, which indicates
the asteroseismic radius correction factors that have been inferred in this work are not determined by choice of color or magnitude terms in the {\it Gaia} parallax zero-point. Regarding the global term, $c = 52.8 \muas$, we show
in \cite{zinn+2019} that the global parallax correction behaves differently than an asteroseismic radius correction factor. In this work, we have been conservative in our approach by inferring radius correction factors using only high-parallax stars ($\varpi > 1\mas$), which are essentially
unaffected by a {\it Gaia} parallax zero-point correction of $\approx 0.05\mas$. Not only should high-parallax stars be unbiased indicators of the radius agreement, but their asteroseismic parallaxes are more sensitive to errors in the asteroseismic radius scale than small-parallax stars \citep{zinn+2019}, and therefore are doubly useful for fitting the radius correction factors ($a_1$-$a_4$ in Equation~\ref{eq:broken}; see \S\ref{sec:abs}). Looking at the
stars least affected by a {\it Gaia} parallax correction in this way, we found absolute agreement between the asteroseismic radius scale and the {\it Gaia} radius scale is within $2\% \pm 2\% {\rm\, (syst.)}$ level for stars with radii below
$R = 30\rsun$. We also examined the differential trends using the full giant sample, which includes stars with small parallax (\S\ref{sec:diff2}). The flat trend with parallax of the radius agreement shown in Figure~\ref{fig:v_plx} demonstrates that even these small-parallax giants have unbiased {\it Gaia} radii following a zero-point correction to the {\it Gaia} parallax scale. If errors in the parallax offset existed at the $\pm 9\muas$ level (the systematic error on the global parallax offset from \cite{zinn+2019}, and which is included in our $2\%$ systematic uncertainty in the radius agreement), they would manifest as trends denoted by the solid grey curves in Figure~\ref{fig:v_plx}.

Regarding the effect of the temperature scale on our results, we quantified the systematic effect of global temperature shifts to be at the $1\%$ level. We illustrate with Figures~\ref{fig:money}a~\&\ref{fig:money}c how the radius agreement changes if the APOGEE temperature scale were smaller by 40K (Figure~\ref{fig:money}a) and larger by 40K (Figure~\ref{fig:money}c). These temperature variations would constitute a $2\sigma$ systematic error according to our systematic uncertainty budget from \S\ref{sec:teff_syst}, and in this sense represent an extreme example of the effect of temperature systematics. In these panels, we have included the effect of a temperature shift on the bolometric correction, which tends to moderate the effect of temperature on the radius, such that the {\it Gaia} radius does not scale as strongly with temperature as Equation~\ref{eq:plx} implies.

We have also verified that systematics due to the choice of $f_{\dnu}$ (which affects the asteroseismic radii according to Equation~\ref{eq:scaling3}) does not significantly impact our results by using the prescription from \cite{sharma+2016} instead of using our nominal BeSPP $f_{\dnu}$ values.\footnote{The \cite{sharma+2016} code for computing $f_{\dnu}$, \texttt{asfgrid} \citep{asfgrid}, is available at \url{http://www.physics.usyd.edu.au/k2gap/Asfgrid/}.}

\section{Conclusions}
\begin{enumerate}
\item For radii between $0.8\rsun$ and $30\rsun$ we conclude that the asteroseismic radius scale and the {\it Gaia} radius scale agree within $2\%$, which is within systematic uncertainties. There appear to be differential trends as a function of radius in this agreement, which are statistically significant ($4\% \pm 0.6\%$).
    \item Our results agree with those from \cite{hall+2019}, who performed a comparison of the asteroseismic and {\it Gaia} red clump absolute luminosity. In that work, the asteroseismic radii of the red clump stars were found to be larger than those from {\it Gaia}, which could be corrected by adjusting the temperature scale by 70K. Here, we find a similar level of radius inflation, but can only attribute $1\%$ of our $2\%$ total systematic uncertainty on the radius inflation to temperature effects, because of the $0.5\%$ accuracy of the infrared flux method temperature calibration. 
  \item After correcting {\it
    Gaia} parallaxes and asteroseismic radii according to our
    best-fitting model, the largest stars in our sample, with $R > 30 \rsun$, have asteroseismic radii that are too large by
$8.7 \pm 0.9 \% {\rm\, (rand.)} \pm 2.0\% {\rm\, (syst.)}$.
  \item We quantify the spatial correlations of {\it Gaia} parallaxes for the {\it Kepler} field, but find they are unimportant for our analysis. At scales of $0.05^{\circ}$, $1^{\circ}$, and $5^{\circ}$, a typical parallax systematic error floor given a statistical uncertainty on parallax of $\sigma_{\varpigaia}$ would be $0.1\sigma_{\varpigaia}$, $0.07\sigma_{\varpigaia}$, and $0.016\sigma_{\varpigaia}$, respectively.
  \item By investigating systematics in our radii due to bolometric
    corrections, we find that reasonable bolometric correction choices
    from the literature disagree at the $2\%$ level, which suggests
    that a percent level fundamental bolometric correction scale is difficult to
    arrive at.
    \item We find only marginal evidence for an asteroseismic radius inflation of $2\% \pm 2\%  {\rm\, (rand.)} \pm 2\% {\rm\, (syst.)}$ and mass inflation of $6\% \pm 8\%  {\rm\, (rand.)} \pm 7\% {\rm\, (syst.)}$ for low-metallicity stars,
[Fe/H] $ < -1.0$. For more solar-like 
metallicities, there are also no significant metallicity-dependent radius
anomalies, to within $0.5\%$ per dex in metallicity for giants and $1.1\%$ per dex for dwarfs/subgiants.
\end{enumerate}
  
In light of the remarkable agreement between asteroseismology and a fundamental parallactic radius scale, the systematics in bolometric correction, extinction, and temperature that we have identified in this work will likely limit future work on constraining the asteroseismic radius scale. For this reason, we are currently investigating the origin of the seemingly inflated asteroseismic radii for the most evolved giants in our sample ($ 30\rsun \leq \rscal < 50 \rsun$), whose scaling relation radii disagree beyond our nominal systematics level of $2\%$. It is likely the case that additional systematics will be significant in this regime (e.g., $\numax$ measurement errors). Nevertheless, we believe that accounting for non-adiabatic effects in pulsation models in evolved stars could help explain the radius inflation we observe in this work, and are thus conducting a complementary theoretical approach to understand these observations.

\nocite{pandas}

\section{Acknowledgments}
M.~H.~P. and J.~Z. acknowledge support from NASA grants 80NSSC18K0391 and
NNX17AJ40G. D.H. acknowledges support by the National Science Foundation (AST-1717000). D.~S. is the recipient of an Australian Research Council Future Fellowship (project number
FT1400147). Parts of this research were conducted by the Australian
Research Council Centre of Excellence for All Sky Astrophysics in 3
Dimensions (ASTRO 3D), through project number CE170100013. A.~S. is partially supported by MICINN grant ESP2017-82674-R and Generalitat de Catalunya grant 2017-SGR-1131. This research was supported in part by the National Science Foundation under Grant No. NSF PHY-1748958 and by the Heising-Simons Foundation.

This paper includes data collected by the {\it Kepler}
mission. Funding for the {\it Kepler} mission is provided by the NASA Science Mission directorate.

This work has made use of data from the European Space Agency (ESA)
mission
{\it Gaia} (\url{https://www.cosmos.esa.int/gaia}), processed by the
{\it Gaia}
Data Processing and Analysis Consortium (DPAC,
\url{https://www.cosmos.esa.int/web/gaia/dpac/consortium}). Funding
for the DPAC
has been provided by national institutions, in particular the
institutions
participating in the {\it Gaia} Multilateral Agreement.

This paper includes data collected by the {\it Kepler}
mission. Funding for the {\it Kepler} mission is provided by the NASA
Science Mission directorate.

This research was partially conducted during the Exostar19 program at the Kavli Institute for Theoretical Physics at UC Santa Barbara, which was supported in part by the National Science Foundation under Grant No. NSF PHY-1748958.

Funding for the Sloan Digital Sky Survey IV has been provided by the Alfred P. Sloan Foundation, the U.S. Department of Energy Office of Science, and the Participating Institutions. SDSS-IV acknowledges
support and resources from the Center for High-Performance Computing at
the University of Utah. The SDSS web site is www.sdss.org.

SDSS-IV is managed by the Astrophysical Research Consortium for the 
Participating Institutions of the SDSS Collaboration including the 
Brazilian Participation Group, the Carnegie Institution for Science, 
Carnegie Mellon University, the Chilean Participation Group, the French Participation Group, Harvard-Smithsonian Center for Astrophysics, 
Instituto de Astrof\'isica de Canarias, The Johns Hopkins University, Kavli Institute for the Physics and Mathematics of the Universe (IPMU) / 
University of Tokyo, the Korean Participation Group, Lawrence Berkeley National Laboratory, 
Leibniz Institut f\"ur Astrophysik Potsdam (AIP),  
Max-Planck-Institut f\"ur Astronomie (MPIA Heidelberg), 
Max-Planck-Institut f\"ur Astrophysik (MPA Garching), 
Max-Planck-Institut f\"ur Extraterrestrische Physik (MPE), 
National Astronomical Observatories of China, New Mexico State University, 
New York University, University of Notre Dame, 
Observat\'ario Nacional / MCTI, The Ohio State University, 
Pennsylvania State University, Shanghai Astronomical Observatory, 
United Kingdom Participation Group,
Universidad Nacional Aut\'onoma de M\'exico, University of Arizona, 
University of Colorado Boulder, University of Oxford, University of Portsmouth, 
University of Utah, University of Virginia, University of Washington, University of Wisconsin, 
Vanderbilt University, and Yale University.

\software{BeSPP \citep{serenelli+2013a,serenelli17}, asfgrid \citep{asfgrid}, emcee \citep{foreman-mackey+2013}, NumPy \citep{numpy}, pandas \citep{pandas}, Matplotlib \citep{matplotlib}, IPython \citep{ipython}}

\clearpage
\newpage

\appendix
\section{Bolometric correction and extinction systematics}
\label{app:fbol}
Our adopted bolometric
scale in this work is the MIST $\ks$-band bolometric correction, $BC_{\ks}$, and
therefore the first test we performed was a self-consistency check of the MIST
bolometric corrections for the giant sample. We started out by assuming extinction
coefficients, $A_{\lambda}/A_{V}$,
for SDSS optical bands, $\lambda = {g, r, i}$ from \cite{an+2009}. We then
derived a visual extinction, $A_V$, based on each SDSS-$\ks$
color. This process of course depends on both the SDSS-band and
$\ks$-band bolometric corrections, and is effectively a test of the
consistency of the bolometric corrections. We compared these
extinctions to a common scale:
our adopted extinction scale from \cite{rodrigues+2014}. We took the
median differences between the SDSS-band MIST extinctions and the
\cite{rodrigues+2014} extinctions for the giant sample as an
indication of the self-consistency of the MIST bolometric
corrections. We found that the $g$-band, $r$-band, and $i$-band MIST extinctions agree with the \cite{rodrigues+2014}
extinctions to within $1.3 \pm 0.3\%$, $3.2 \pm 0.2\%$, and $0.4 \pm
0.4\%$, where the systematic error due to the uncertainty in the
extinction coefficients dominates over the random uncertainty on
the median of the MIST extinctions for the giant sample. We conclude
that the MIST bolometric corrections are consistent with each other to at least $3\%$.

Ultimately, the quantity that we would like to pin down is not the the $\ks$-band bolometric correction, but rather the
bolometric flux itself. This quantity of course depends on not only the bolometric
correction, but also the adopted extinction. We have adopted an
infrared-based bolometric flux because of the relative insensitivity
to extinction. Using the bolometric correction, we de-extinct the 2MASS $\ks$ photometry by converting our $A_V$ from \cite{rodrigues+2014} into $A_{\ks}$ by way of an infrared extinction coefficient, as mentioned in \S\ref{sec:gaia}. We adopt a solar irradiance from \cite{mamajek+2015a}, $f_0 = 1.361\times10^{6} erg/s/cm^2$, and assume an apparent bolometric magnitude of $m_{\mathrm{cal}} = -26.82$ (using the visual magnitude of the Sun, $V_{\odot} = -26.76$, and its visual bolometric correction, $BC_{V,\odot} = -0.06$; \citealt{torres+2010a}). The bolometric flux is then $f_{\mathrm{bol}} =  f_0 10^{-0.4(K_{\mathrm{s}} -m_{\mathrm{cal}} + BC - A_{K\mathrm{s}})}$. To test the accuracy of our MIST $\ks$ bolometric flux scale, we have computed bolometric fluxes for comparison using several other approaches, which are described below.

First, we compare to a bolometric flux computed via spectral energy
distribution (SED) fitting described in the main text. We computed the bolometric fluxes using this method for all
giant stars with positive parallax and parallax errors less than
$20\%$. The SED fitting was initialized with an initial guess for the
extinction taken to be the \cite{rodrigues+2014}
extinction. 

We also compare the bolometric fluxes we use to those from the IRFM method described in the main text. The IRFM hinges on a different
dependence on temperature of the visual and infrared flux to
iteratively estimate temperature and angular diameter (and bolometric
flux). As the name implies, this method
requires infrared photometry, for which we use $J$, $H$, and
$K_{\mathrm{s}}$ from 2MASS. By way
of visual photometry, we used $g$ and $r$ photometry from the {\it Kepler} Input Catalogue
\citep[KIC;][]{brown+2011}, which has been re-calibrated to be on the Sloan Digital Sky
Survey \citep[SDSS;][]{abolfathi+2018a} scale
by \cite{pinsonneault+2012a}. As
implemented in \cite{ghb09}, the IRFM requires $V$-band photometry,
and so we transform $g$ and $r$ magnitudes to Johnson $B$
and $V$ according to Lupton
(2005)\footnote{\url{https://www.sdss3.org/dr10/algorithms/sdssUBVRITransform.php}}. The
extinctions in the de-extinction procedure are our adopted
\cite{rodrigues+2014} extinctions.

\begin{figure}[h!]
\includegraphics[width=0.5\textwidth]{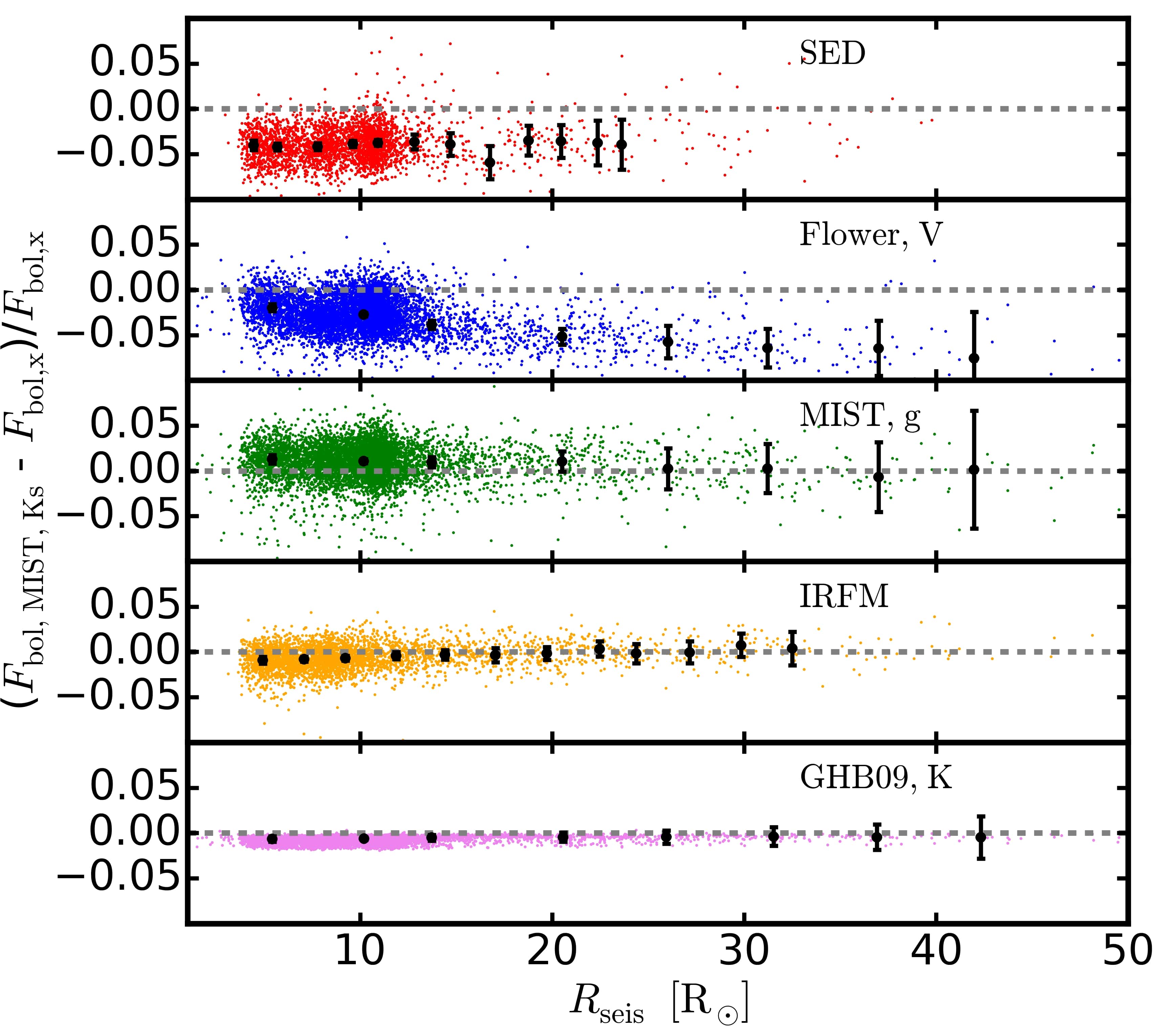} 
\caption{Fractional difference in our adopted $\ks$-band bolometric
  fluxes computed using MIST bolometric corrections and
  extinctions from \cite{rodrigues+2014} and various other bolometric
  flux systems, as a function of radius. See text for details.}
\label{fig:fbol}
\end{figure}

\begin{figure}[h!]
\includegraphics[width=0.5\textwidth]{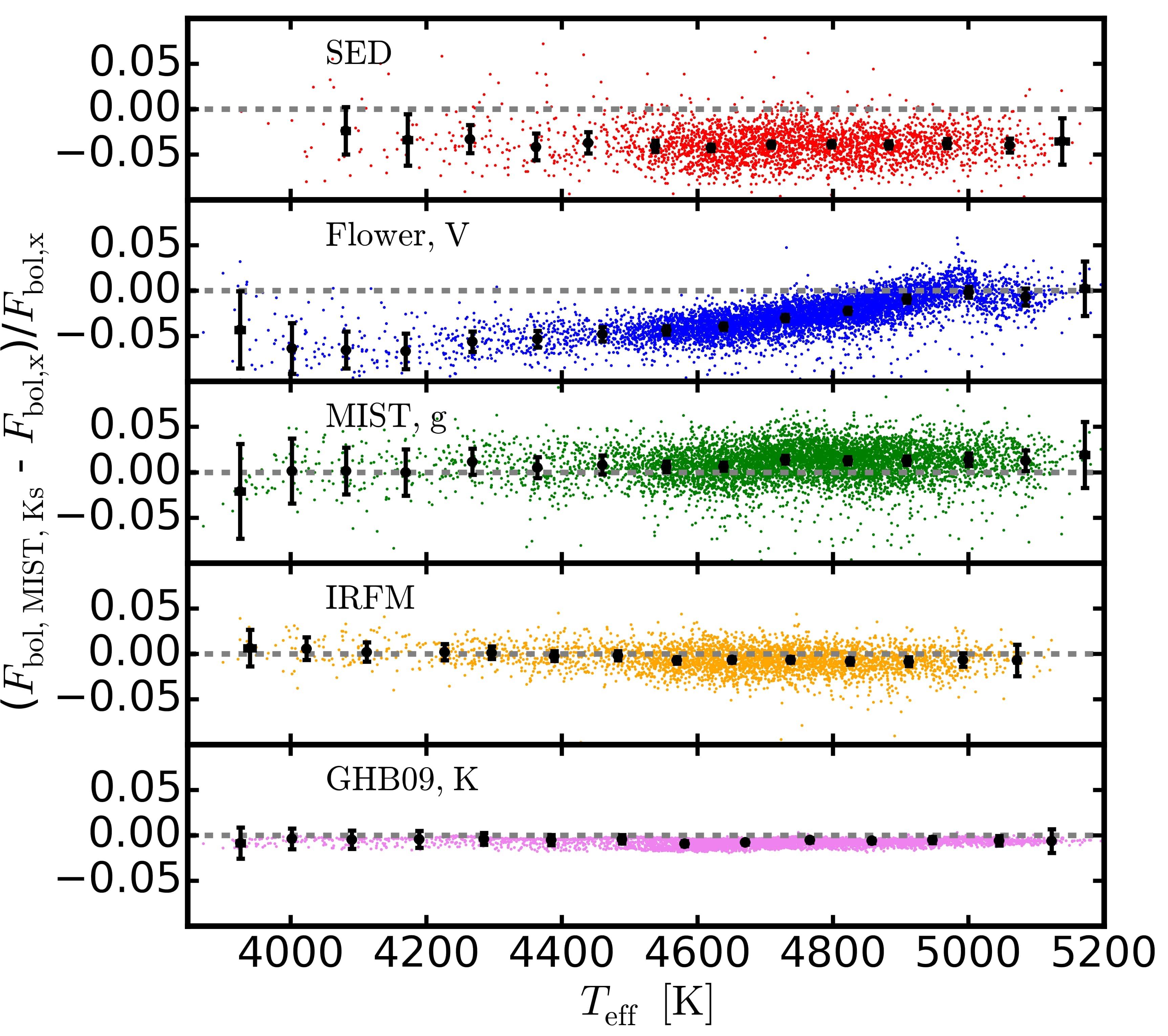} 
\caption{The same as Figure~\protect\ref{fig:fbol}, except plotted as a
  function of temperature.}
\label{fig:fbol_teff}
\end{figure}

The SED and IRFM bolometric fluxes are compared to our
adopted $\ks$-band MIST bolometric fluxes in
Figure~\ref{fig:fbol}. Also shown are three more sets of bolometric fluxes computed assuming the \cite{rodrigues+2014}
extinctions: one using a $g$-band MIST bolometric correction; another the empirical visual bolometric correction from
\cite{flower+1996}; and another using the $\ks$-band bolometric
correction from \cite{ghb09} (``GHB09, K" in the figure).

The figure demonstrates first and foremost that the agreement across
these methods is globally good. This is especially true when considering that the bolometric corrections span a two-decade range in publication date: from 1996 to present. In particular, this figure demonstrates excellent
agreement in the mean fluxes ($0.73 \pm 0.09\%$) between our adopted $\ks$-band MIST bolometric fluxes and
the $\ks$-band bolometric fluxes using the bolometric correction from \cite{ghb09}. Part of this agreement is certainly due to the fact that any infrared flux scale is insensitive to extinction choice, but it more importantly establishes a consensus in the infrared bolometric corrections. Indeed, there
is also excellent agreement with the IRFM bolometric flux scale ($0.66
\pm 0.11\%$). This, even though the IRFM scale
incorporates visual information ($B$ and $V$), and therefore depends to some extent on the \cite{rodrigues+2014} extinctions.

The largest deviations in bolometric flux scale are between $\ks$ MIST \& SED (mean difference of $3.8 \pm 0.1\%$) and between $\ks$ MIST \& $V$-band ($3.0 \pm 0.1\%$). As we see in Figure~\ref{fig:fbol_teff}, the disagreement between our adopted infrared scale and the $V$-band scale is a strong function of temperature, which suggests there are genuine disagreements between the MIST models and the empirical $V$-band bolometric corrections. 
Unlike the other approaches, the SED approach does not assume the \cite{rodrigues+2014} extinctions. Differences in model atmospheres between those used in the C3K grid (Conroy
et al., in prep) and those used in the SED approach described in \cite{stassun+2016a} and \cite{stassun+2017a} would result in different extinctions and bolometric corrections, both of which would affect bolometric flux agreement. On the extinction side, the predicted extinctions using the SED approach differ by $\sim$ 0.2mag from the extinctions from \cite{rodrigues+2014}. If adopting the bolometric fluxes from \cite{rodrigues+2014} and not allowing extinction as a free parameter in the SED fitting process, the SED bolometric fluxes would shift to be about $3\%$ {\it lower} compared to our adopted $\ks$-band bolometric fluxes (otherwise, they sit at about $4\%$ {\it higher} than the infrared fluxes). Shifts in extinction estimates from the SED fitting approach, in other words, map to shifts in bolometric fluxes. Given the relative insensitivity of the infrared bolometric fluxes to the choice of extinction, there are likely model color differences among \cite{rodrigues+2014}, \cite{stassun+2016a}, and C3K that would explain both 1) the different extinctions from the SED approach of \cite{stassun+2016a} and from that of \cite{rodrigues+2014} and 2) the remaining $3\%$ difference between the SED and the MIST $\ks$-band bolometric fluxes when fixing the SED extinctions to those from \cite{rodrigues+2014}.

The bolometric corrections we have discussed here reflect substantive differences in approach, as well as choice in adopted atmosphere models. 
For these reasons, we interpret these differences in the bolometric flux scale as $2\sigma$ systematics. So while on the face of it, the largest mean offset in the bolometric corrections is $\sim 4\%$, we adopt this as a $2\%$ systematic at the $1\sigma$ level. This choice for the systematic uncertainty in the bolometric correction scale for our work reflects the understanding, for instance, that the underlying atmosphere models for these two bolometric corrections (C3K and SED) are separated by 26 years, and have significant departures in, e.g., adopted line lists. Ultimately, the largest differences we note in bolometric flux ($\sim 2-4\%$) map to differences of $1-2\%$ in radius space, as Table~\ref{tab:results} indicates. 

\section{Spatial correlations in DR2 parallaxes}
\label{app:cov}
Having corrected for global, color-, and magnitude-dependent terms in the zero-point in {\it Gaia} parallaxes, we need to similarly account for the spatial dependence in the zero-point. The effect of spatial correlations in parallax can inflate the random error on inferred quantities in our sample, and so we describe here how we go about quantifying the off-diagonal elements in the covariance matrix, $C$.

\cite{zinn+2019} quantified the spatial-dependence of the offset
between parallaxes derived from asteroseismology (calculated according
to Equation~\ref{eq:plx}) and those from \textit{Gaia} DR1. The basis of
the inference of spatially-correlated systematics was a Pearson correlation coefficient that
described the correlation between the quantity $\varpigaia - \varpiast$
as a function of angular separation on the sky. This correlation
function would be positive when two regions of the sky
separated by an angular distance, $\Delta \theta$, had a \textit{Gaia}
parallax measurement that were both too low or both too high compared
to the asteroseismic parallax, indicating a positive correlation at a
certain angular scale. A negative angular correlation would exist
where two patches of sky had \textit{Gaia} parallaxes that were offset
from the asteroseismic parallaxes in opposite directions. Where the two parallaxes agreed, the quantity
would be zero. 

We compute the binned Pearson correlation coefficient, correcting the {\it Gaia} parallaxes
according to the zero-point model from \cite{zinn+2019} using the full giant sample, and then also remove any residual median in the difference in parallax scales. (If we were not to correct the
{\it Gaia} parallaxes for global, magnitude-, and color-dependent errors
before fitting for the spatial correlations, we would find a
too-large spatial parallax correlation due to the global offset
between asteroseismic and {\it Gaia} parallaxes across the entire {\it
  Kepler} field.)
  
\begin{figure}
\includegraphics[width=0.5\textwidth]{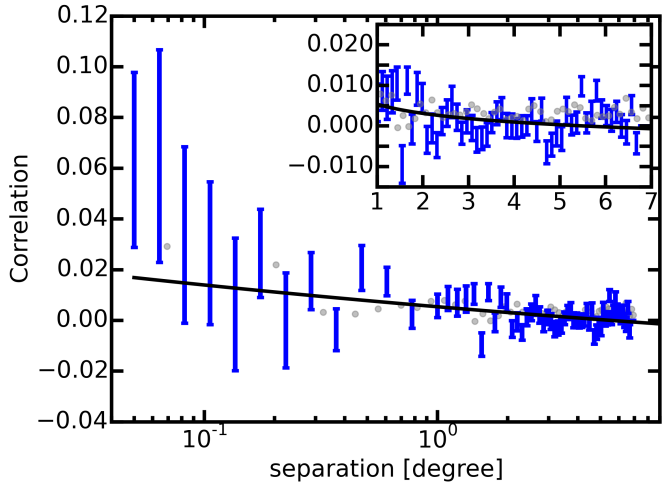}
\caption{Error bars show the binned Pearson correlation coefficient of the
asteroseismic-{\rm Gaia} parallax difference as a function of angular
separation. The black curve shows the fit using Equation~\ref{eq:corr}. The points are spatial covariance points from
the bottom panel of L18's Figure 14, re-scaled to be a binned
correlation coefficient by assuming a typical error for their quasar
sample of $0.25$mas.}
\label{fig:corr}
\end{figure}

We fit the correlation coefficient of the parallax difference as a
function of angular separation on the sky, $\Delta \theta$, with the following
model:

\begin{align}
  \begin{split}
    \chi(\Delta \theta) &= H(\Delta \theta)[ A \exp{(- \ln 2 \ln \Delta \theta /\theta_{1/2})} + B ]
  \end{split}
\label{eq:corr}
\end{align}
where $A$ is a characteristic amplitude to the correlations; $\theta_{1/2}$ is a characteristic angular scale; and $B$ is a constant.  The Heaviside function, $H(\Delta \theta)$, ensures that the correlation is set to zero for
the same star $\chi(\Delta \theta = 0) = 0$. We follow the approach described in \cite{zinn+2019} to fit this functional form to the binned Pearson correlation coefficient. In this approach, the correlations between adjacent bins in the Pearson correlation coefficient (error bars in Figure~\ref{fig:corr}) are taken into account, and the model is fitted using MCMC. We do not take into account edge effects as \cite{zinn+2019} do by fitting to simulated data.
The best-fitting parameters for Equation~\ref{eq:corr} and their $1\sigma$ uncertainties
are given in Table \ref{tab:corr}.

The observed correlation coefficient for our sample, along with the
best-fitting model from Equation~\ref{eq:corr} is shown in
Figure~\ref{fig:corr}. We use this model for the angular parallax correlation, $\chi(\Delta \theta)$, in our covariance matrix
when taking into account spatial correlations in parallax (Equation~\ref{eq:second_model}). According to this best-fitting model, the level of correlation at angular separations of $0.05^{\circ}$ is 0.02, and decreases to 0.01 at $1^{\circ}$, and is 0.0003 at $5^{\circ}$. This means one cannot reduce the parallax uncertainty when averaging over more than 60, 200, or 4000 stars at these angular separations.

We find that our covariance agrees well with the covariance reported
by \citep[][; L18]{lindegren+2018} under a simple re-scaling, assuming the median error of their
quasar sample is $0.25$mas. We show the resulting data points from
L18's Figure 14 in our Figure~\ref{fig:corr}. The exponential
behavior at $\Delta \theta \lesssim 0.1^{\circ}$ is similar to ours, and both our and L18's measurements indicate the presence of small-amplitude
oscillatory behavior.

Whether or not we include the full covariance matrix in our analysis, according to Equations~\ref{eq:cov}~\&~\ref{eq:second_model}, our results are unaffected (compare ``K MIST no cov" and ``K MIST" entries in Table~\ref{tab:results}). This can be understood by the fact that the variability in the {\it Gaia} parallax scale as a function of position averages out over the {\it Kepler} field of view, leaving unaffected the central values of our radius agreement fit. Moreover, the relatively small number of stars in this high-parallax sub-sample means that one does not average down by $1/\sqrt{N}$ to the systematic floor set by the spatial correlations.

\begin{table*}
  \begin{tabular*}{\textwidth}{ccccccccc}
  $A$ & $\theta_{1/2}$ & $B$ & $\chi^2/dof$ \\ \hline 
$4.031 \times 10^{-2} \pm 5.796 \times 10^{-5}$ &  $8.3 \pm  2.3^{\circ}$ & $-3.497 \times 10^{-2} \pm 5.604 \times 10^{-5}$ & 7.930\\ \hline
\end{tabular*}
\caption{The best-fitting parameters for Equation~\ref{eq:corr}.}
\label{tab:corr}
\end{table*}

\clearpage
\newpage

\bibliography{zinn_radius_arxiv}

\begin{thebibliography}{}
\expandafter\ifx\csname natexlab\endcsname\relax\def\natexlab#1{#1}\fi
\providecommand{\url}[1]{\href{#1}{#1}}
\providecommand{\dodoi}[1]{doi:~\href{http://doi.org/#1}{\nolinkurl{#1}}}
\providecommand{\doeprint}[1]{\href{http://ascl.net/#1}{\nolinkurl{http://ascl.net/#1}}}
\providecommand{\doarXiv}[1]{\href{https://arxiv.org/abs/#1}{\nolinkurl{https://arxiv.org/abs/#1}}}

\bibitem[{{Abolfathi} {et~al.}(2018){Abolfathi}, {Aguado}, {Aguilar}, {Allende
  Prieto}, {Almeida}, {Ananna}, {Anders}, {Anderson}, {Andrews}, {Anguiano}, \&
  et~al.}]{abolfathi+2018a}
{Abolfathi}, B., {Aguado}, D.~S., {Aguilar}, G., {et~al.} 2018, \apjs, 235, 42

\bibitem[{{Alonso} {et~al.}(1994){Alonso}, {Arribas}, \&
  {Martinez-Roger}}]{alonso+1994a}
{Alonso}, A., {Arribas}, S., \& {Martinez-Roger}, C. 1994, \aap, 282, 684

\bibitem[{{Alonso} {et~al.}(1999){Alonso}, {Arribas}, \&
  {Mart{\'{\i}}nez-Roger}}]{alonso+1999a}
{Alonso}, A., {Arribas}, S., \& {Mart{\'{\i}}nez-Roger}, C. 1999, \aaps, 139,
  335, \dodoi{10.1051/aas:1999506}

\bibitem[{{An} {et~al.}(2019){An}, {Pinsonneault}, {Terndrup}, \&
  {Chung}}]{an+2019}
{An}, D., {Pinsonneault}, M.~H., {Terndrup}, D.~M., \& {Chung}, C. 2019, \apj,
  879, 81, \dodoi{10.3847/1538-4357/ab23ed}

\bibitem[{{An} {et~al.}(2009){An}, {Pinsonneault}, {Masseron}, {Delahaye},
  {Johnson}, {Terndrup}, {Beers}, {Ivans}, \& {Ivezi{\'c}}}]{an+2009}
{An}, D., {Pinsonneault}, M.~H., {Masseron}, T., {et~al.} 2009, \apj, 700, 523,
  \dodoi{10.1088/0004-637X/700/1/523}

\bibitem[{{Arribas} \& {Martinez Roger}(1987)}]{arribas+1987a}
{Arribas}, S., \& {Martinez Roger}, C. 1987, \aap, 178, 106

\bibitem[{{Bailer-Jones} {et~al.}(2018){Bailer-Jones}, {Rybizki}, {Fouesneau},
  {Mantelet}, \& {Andrae}}]{bailer-jones+2018a}
{Bailer-Jones}, C.~A.~L., {Rybizki}, J., {Fouesneau}, M., {Mantelet}, G., \&
  {Andrae}, R. 2018, \aj, 156, 58

\bibitem[{{Ball} \& {Gizon}(2014)}]{ball_gizon_2014}
{Ball}, W.~H., \& {Gizon}, L. 2014, \aap, 568, A123

\bibitem[{{Bertelli} {et~al.}(1994){Bertelli}, {Bressan}, {Chiosi}, {Fagotto},
  \& {Nasi}}]{bertelli+1994a}
{Bertelli}, G., {Bressan}, A., {Chiosi}, C., {Fagotto}, F., \& {Nasi}, E. 1994,
  \aaps, 106, 275

\bibitem[{{Blackwell} {et~al.}(1980){Blackwell}, {Petford}, \&
  {Shallis}}]{blackwell+1980a}
{Blackwell}, D.~E., {Petford}, A.~D., \& {Shallis}, M.~J. 1980, \aap, 82, 249

\bibitem[{{Blackwell} \& {Shallis}(1977)}]{blackwell+1977a}
{Blackwell}, D.~E., \& {Shallis}, M.~J. 1977, \mnras, 180, 177

\bibitem[{{Bovy} {et~al.}(2016){Bovy}, {Rix}, {Green}, {Schlafly}, \&
  {Finkbeiner}}]{bovy+2016}
{Bovy}, J., {Rix}, H.-W., {Green}, G.~M., {Schlafly}, E.~F., \& {Finkbeiner},
  D.~P. 2016, \apj, 818, 130, \dodoi{10.3847/0004-637X/818/2/130}

\bibitem[{{Brogaard} {et~al.}(2018){Brogaard}, {Hansen}, {Miglio}, {Slumstrup},
  {Frandsen}, {Jessen-Hansen}, {Lund}, {Bossini}, {Thygesen}, {Davies},
  {Chaplin}, {Arentoft}, {Bruntt}, {Grundahl}, \& {Handberg}}]{brogaard+2018}
{Brogaard}, K., {Hansen}, C.~J., {Miglio}, A., {et~al.} 2018, \mnras, 476,
  3729, \dodoi{10.1093/mnras/sty268}

\bibitem[{{Brown} {et~al.}(1991){Brown}, {Gilliland}, {Noyes}, \&
  {Ramsey}}]{brown+1991}
{Brown}, T.~M., {Gilliland}, R.~L., {Noyes}, R.~W., \& {Ramsey}, L.~W. 1991,
  \apj, 368, 599, \dodoi{10.1086/169725}

\bibitem[{{Brown} {et~al.}(2011){Brown}, {Latham}, {Everett}, \&
  {Esquerdo}}]{brown+2011}
{Brown}, T.~M., {Latham}, D.~W., {Everett}, M.~E., \& {Esquerdo}, G.~A. 2011,
  \aj, 142, 112, \dodoi{10.1088/0004-6256/142/4/112}

\bibitem[{{Casagrande} {et~al.}(2010){Casagrande}, {Ram{\'{\i}}rez},
  {Mel{\'e}ndez}, {Bessell}, \& {Asplund}}]{casagrande+2010}
{Casagrande}, L., {Ram{\'{\i}}rez}, I., {Mel{\'e}ndez}, J., {Bessell}, M., \&
  {Asplund}, M. 2010, \aap, 512, A54

\bibitem[{{Chaplin} {et~al.}(2014){Chaplin}, {Elsworth}, {Davies}, {Campante},
  {Handberg}, {Miglio}, \& {Basu}}]{chaplin14}
{Chaplin}, W.~J., {Elsworth}, Y., {Davies}, G.~R., {et~al.} 2014, \mnras, 445,
  946, \dodoi{10.1093/mnras/stu1811}

\bibitem[{{Chaplin} {et~al.}(2011){Chaplin}, {Kjeldsen},
  {Christensen-Dalsgaard}, {Basu}, {Miglio}, {Appourchaux}, {Bedding},
  {Elsworth}, {Garc{\'{\i}}a}, {Gilliland}, {Girardi}, {Houdek}, {Karoff},
  {Kawaler}, {Metcalfe}, {Molenda-{\.Z}akowicz}, {Monteiro}, {Thompson},
  {Verner}, {Ballot}, {Bonanno}, {Brand{\~a}o}, {Broomhall}, {Bruntt},
  {Campante}, {Corsaro}, {Creevey}, {Do{\u g}an}, {Esch}, {Gai}, {Gaulme},
  {Hale}, {Handberg}, {Hekker}, {Huber}, {Jim{\'e}nez}, {Mathur}, {Mazumdar},
  {Mosser}, {New}, {Pinsonneault}, {Pricopi}, {Quirion}, {R{\'e}gulo},
  {Salabert}, {Serenelli}, {Silva Aguirre}, {Sousa}, {Stello}, {Stevens},
  {Suran}, {Uytterhoeven}, {White}, {Borucki}, {Brown}, {Jenkins}, {Kinemuchi},
  {Van Cleve}, \& {Klaus}}]{chaplin+2011c}
{Chaplin}, W.~J., {Kjeldsen}, H., {Christensen-Dalsgaard}, J., {et~al.} 2011,
  Science, 332, 213

\bibitem[{{Choi} {et~al.}(2016){Choi}, {Dotter}, {Conroy}, {Cantiello},
  {Paxton}, \& {Johnson}}]{choi+2016a}
{Choi}, J., {Dotter}, A., {Conroy}, C., {et~al.} 2016, \apj, 823, 102

\bibitem[{{Christensen-Dalsgaard}(1993)}]{christensendalsgaard1993}
{Christensen-Dalsgaard}, J. 1993, in Astronomical Society of the Pacific
  Conference Series, Vol.~42, GONG 1992. Seismic Investigation of the Sun and
  Stars, ed. T.~M. {Brown}, 347

\bibitem[{{Dotter}(2016)}]{dotter+2016a}
{Dotter}, A. 2016, \apjs, 222, 8

\bibitem[{{Epstein} {et~al.}(2014){Epstein}, {Elsworth}, {Johnson}, {Shetrone},
  {Mosser}, {Hekker}, {Tayar}, {Harding}, {Pinsonneault}, {Silva Aguirre},
  {Basu}, {Beers}, {Bizyaev}, {Bedding}, {Chaplin}, {Frinchaboy},
  {Garc{\'\i}a}, {Garc{\'\i}a P{\'e}rez}, {Hearty}, {Huber}, {Ivans},
  {Majewski}, {Mathur}, {Nidever}, {Serenelli}, {Schiavon}, {Schneider},
  {Sch{\"o}nrich}, {Sobeck}, {Stassun}, {Stello}, \& {Zasowski}}]{epstein+2014}
{Epstein}, C.~R., {Elsworth}, Y.~P., {Johnson}, J.~A., {et~al.} 2014, \apj,
  785, L28, \dodoi{10.1088/2041-8205/785/2/L28}

\bibitem[{{Fitzpatrick}(1999)}]{fitzpatrick1999}
{Fitzpatrick}, E.~L. 1999, \pasp, 111, 63, \dodoi{10.1086/316293}

\bibitem[{{Flower}(1996)}]{flower+1996}
{Flower}, P.~J. 1996, \apj, 469, 355, \dodoi{10.1086/177785}

\bibitem[{{Foreman-Mackey} {et~al.}(2013){Foreman-Mackey}, {Hogg}, {Lang}, \&
  {Goodman}}]{foreman-mackey+2013}
{Foreman-Mackey}, D., {Hogg}, D.~W., {Lang}, D., \& {Goodman}, J. 2013, \pasp,
  125, 306, \dodoi{10.1086/670067}

\bibitem[{{Frandsen} {et~al.}(2013){Frandsen}, {Lehmann}, {Hekker},
  {Southworth}, {Debosscher}, {Beck}, {Hartmann}, {Pigulski}, {Kopacki},
  {Ko{\l}aczkowski}, {St{\c e}{\'s}licki}, {Thygesen}, {Brogaard}, \&
  {Elsworth}}]{frandsen+2013}
{Frandsen}, S., {Lehmann}, H., {Hekker}, S., {et~al.} 2013, \aap, 556, A138,
  \dodoi{10.1051/0004-6361/201321817}

\bibitem[{{Gaia Collaboration} {et~al.}(2016){Gaia Collaboration}, {Prusti},
  {de Bruijne}, {Brown}, {Vallenari}, {Babusiaux}, {Bailer-Jones}, {Bastian},
  {Biermann}, {Evans}, \& et~al.}]{gaia2016}
{Gaia Collaboration}, {Prusti}, T., {de Bruijne}, J.~H.~J., {et~al.} 2016,
  \aap, 595, A1, \dodoi{10.1051/0004-6361/201629272}

\bibitem[{{Gaia Collaboration} {et~al.}(2018){Gaia Collaboration}, {Brown},
  {Vallenari}, {Prusti}, {de Bruijne}, {Babusiaux}, {Bailer-Jones}, {Biermann},
  {Evans}, {Eyer}, \& et~al.}]{gaia2018}
{Gaia Collaboration}, {Brown}, A.~G.~A., {Vallenari}, A., {et~al.} 2018, \aap,
  616, A1, \dodoi{10.1051/0004-6361/201833051}

\bibitem[{{Gaulme} {et~al.}(2016){Gaulme}, {McKeever}, {Jackiewicz}, {Rawls},
  {Corsaro}, {Mosser}, {Southworth}, {Mahadevan}, {Bender}, \&
  {Deshpande}}]{gaulme+2016}
{Gaulme}, P., {McKeever}, J., {Jackiewicz}, J., {et~al.} 2016, \apj, 832, 121,
  \dodoi{10.3847/0004-637X/832/2/121}

\bibitem[{{Gonz{\'a}lez Hern{\'a}ndez} \& {Bonifacio}(2009)}]{ghb09}
{Gonz{\'a}lez Hern{\'a}ndez}, J.~I., \& {Bonifacio}, P. 2009, \aap, 497, 497,
  \dodoi{10.1051/0004-6361/200810904}

\bibitem[{{Green} {et~al.}(2015){Green}, {Schlafly}, {Finkbeiner}, {Rix},
  {Martin}, {Burgett}, {Draper}, {Flewelling}, {Hodapp}, {Kaiser}, {Kudritzki},
  {Magnier}, {Metcalfe}, {Price}, {Tonry}, \& {Wainscoat}}]{green+2015}
{Green}, G.~M., {Schlafly}, E.~F., {Finkbeiner}, D.~P., {et~al.} 2015, \apj,
  810, 25, \dodoi{10.1088/0004-637X/810/1/25}

\bibitem[{{Guggenberger} {et~al.}(2016){Guggenberger}, {Hekker}, {Basu}, \&
  {Bellinger}}]{guggenberger+2016}
{Guggenberger}, E., {Hekker}, S., {Basu}, S., \& {Bellinger}, E. 2016, \mnras,
  460, 4277, \dodoi{10.1093/mnras/stw1326}

\bibitem[{{Gunn} {et~al.}(2006){Gunn}, {Siegmund}, {Mannery}, {Owen}, {Hull},
  {Leger}, {Carey}, {Knapp}, {York}, {Boroski}, {Kent}, {Lupton}, {Rockosi},
  {Evans}, {Waddell}, {Anderson}, {Annis}, {Barentine}, {Bartoszek}, {Bastian},
  {Bracker}, {Brewington}, {Briegel}, {Brinkmann}, {Brown}, {Carr},
  {Czarapata}, {Drennan}, {Dombeck}, {Federwitz}, {Gillespie}, {Gonzales},
  {Hansen}, {Harvanek}, {Hayes}, {Jordan}, {Kinney}, {Klaene}, {Kleinman},
  {Kron}, {Kresinski}, {Lee}, {Limmongkol}, {Lindenmeyer}, {Long}, {Loomis},
  {McGehee}, {Mantsch}, {Neilsen}, {Neswold}, {Newman}, {Nitta}, {Peoples},
  {Pier}, {Prieto}, {Prosapio}, {Rivetta}, {Schneider}, {Snedden}, \&
  {Wang}}]{gunn+2006}
{Gunn}, J.~E., {Siegmund}, W.~A., {Mannery}, E.~J., {et~al.} 2006, \aj, 131,
  2332, \dodoi{10.1086/500975}

\bibitem[{{Hall} {et~al.}(2019){Hall}, {Davies}, {Elsworth}, {Miglio},
  {Bedding}, {Brown}, {Khan}, {Hawkins}, {Garc{\'{\i}}a}, {Chaplin}, \&
  {North}}]{hall+2019}
{Hall}, O.~J., {Davies}, G.~R., {Elsworth}, Y.~P., {et~al.} 2019, \mnras, 486,
  3569, \dodoi{10.1093/mnras/stz1092}

\bibitem[{{Holtzman} {et~al.}(2015){Holtzman}, {Shetrone}, {Johnson}, {Allende
  Prieto}, {Anders}, {Andrews}, {Beers}, {Bizyaev}, {Blanton}, {Bovy},
  {Carrera}, {Chojnowski}, {Cunha}, {Eisenstein}, {Feuillet}, {Frinchaboy},
  {Galbraith-Frew}, {Garc{\'{\i}}a P{\'e}rez}, {Garc{\'{\i}}a-Hern{\'a}ndez},
  {Hasselquist}, {Hayden}, {Hearty}, {Ivans}, {Majewski}, {Martell},
  {Meszaros}, {Muna}, {Nidever}, {Nguyen}, {O'Connell}, {Pan}, {Pinsonneault},
  {Robin}, {Schiavon}, {Shane}, {Sobeck}, {Smith}, {Troup}, {Weinberg},
  {Wilson}, {Wood-Vasey}, {Zamora}, \& {Zasowski}}]{holtzman+2015}
{Holtzman}, J.~A., {Shetrone}, M., {Johnson}, J.~A., {et~al.} 2015, \aj, 150,
  148, \dodoi{10.1088/0004-6256/150/5/148}

\bibitem[{{Holtzman} {et~al.}(2018){Holtzman}, {Hasselquist}, {Shetrone},
  {Cunha}, {Allende Prieto}, {Anguiano}, {Bizyaev}, {Bovy}, {Casey},
  {Edvardsson}, {Johnson}, {J{\"o}nsson}, {Meszaros}, {Smith}, {Sobeck},
  {Zamora}, {Chojnowski}, {Fernandez-Trincado}, {Garcia-Hernandez}, {Majewski},
  {Pinsonneault}, {Souto}, {Stringfellow}, {Tayar}, {Troup}, \&
  {Zasowski}}]{holtzman+2018}
{Holtzman}, J.~A., {Hasselquist}, S., {Shetrone}, M., {et~al.} 2018, \aj, 156,
  125, \dodoi{10.3847/1538-3881/aad4f9}

\bibitem[{{Huber} {et~al.}(2009){Huber}, {Stello}, {Bedding}, {Chaplin},
  {Arentoft}, {Quirion}, \& {Kjeldsen}}]{huber+2009}
{Huber}, D., {Stello}, D., {Bedding}, T.~R., {et~al.} 2009, Communications in
  Asteroseismology, 160, 74.
\newblock \doarXiv{0910.2764}

\bibitem[{{Huber} {et~al.}(2012{\natexlab{a}}){Huber}, {Ireland}, {Bedding},
  {Brand{\~a}o}, {Piau}, {Maestro}, {White}, {Bruntt}, {Casagrande},
  {Molenda-{\.Z}akowicz}, {Silva Aguirre}, {Sousa}, {Barclay}, {Burke},
  {Chaplin}, {Christensen-Dalsgaard}, {Cunha}, {De Ridder}, {Farrington},
  {Frasca}, {Garc{\'{\i}}a}, {Gilliland}, {Goldfinger}, {Hekker}, {Kawaler},
  {Kjeldsen}, {McAlister}, {Metcalfe}, {Miglio}, {Monteiro}, {Pinsonneault},
  {Schaefer}, {Stello}, {Stumpe}, {Sturmann}, {Sturmann}, {ten Brummelaar},
  {Thompson}, {Turner}, \& {Uytterhoeven}}]{huber+2012}
{Huber}, D., {Ireland}, M.~J., {Bedding}, T.~R., {et~al.} 2012{\natexlab{a}},
  \apj, 760, 32, \dodoi{10.1088/0004-637X/760/1/32}

\bibitem[{{Huber} {et~al.}(2012{\natexlab{b}}){Huber}, {Ireland}, {Bedding},
  {Howell}, {Maestro}, {M{\'e}rand}, {Tuthill}, {White}, {Farrington},
  {Goldfinger}, {McAlister}, {Schaefer}, {Sturmann}, {Sturmann}, {ten
  Brummelaar}, \& {Turner}}]{huber12}
---. 2012{\natexlab{b}}, \mnras, 423, L16,
  \dodoi{10.1111/j.1745-3933.2012.01242.x}

\bibitem[{{Huber} {et~al.}(2013){Huber}, {Chaplin}, {Christensen-Dalsgaard},
  {Gilliland}, {Kjeldsen}, {Buchhave}, {Fischer}, {Lissauer}, {Rowe},
  {Sanchis-Ojeda}, {Basu}, {Handberg}, {Hekker}, {Howard}, {Isaacson},
  {Karoff}, {Latham}, {Lund}, {Lundkvist}, {Marcy}, {Miglio}, {Silva Aguirre},
  {Stello}, {Arentoft}, {Barclay}, {Bedding}, {Burke}, {Christiansen},
  {Elsworth}, {Haas}, {Kawaler}, {Metcalfe}, {Mullally}, \&
  {Thompson}}]{huber+2013}
{Huber}, D., {Chaplin}, W.~J., {Christensen-Dalsgaard}, J., {et~al.} 2013,
  \apj, 767, 127, \dodoi{10.1088/0004-637X/767/2/127}

\bibitem[{{Huber} {et~al.}(2017){Huber}, {Zinn}, {Bojsen-Hansen},
  {Pinsonneault}, {Sahlholdt}, {Serenelli}, {Silva Aguirre}, {Stassun},
  {Stello}, {Tayar}, {Bastien}, {Bedding}, {Buchhave}, {Chaplin}, {Davies},
  {Garc{\'{\i}}a}, {Latham}, {Mathur}, {Mosser}, \& {Sharma}}]{huber+2017}
{Huber}, D., {Zinn}, J., {Bojsen-Hansen}, M., {et~al.} 2017, \apj, 844, 102,
  \dodoi{10.3847/1538-4357/aa75ca}

\bibitem[{Hunter(2007)}]{matplotlib}
Hunter, J.~D. 2007, Computing in Science \& Engineering, 9, 90,
  \dodoi{10.1109/MCSE.2007.55}

\bibitem[{{Johnson} {et~al.}(2014){Johnson}, {Huber}, {Boyajian}, {Brewer},
  {White}, {von Braun}, {Maestro}, {Stello}, \& {Barclay}}]{johnson14}
{Johnson}, J.~A., {Huber}, D., {Boyajian}, T., {et~al.} 2014, \apj, 794, 15,
  \dodoi{10.1088/0004-637X/794/1/15}

\bibitem[{{Khan} {et~al.}(2019){Khan}, {Miglio}, {Mosser}, {Arenou},
  {Belkacem}, {Brown}, {Katz}, {Casagrand e}, {Chaplin}, {Davies}, {Rendle},
  {Rodrigues}, {Bossini}, {Cantat-Gaudin}, {Elsworth}, {Girardi}, {North}, \&
  {Vallenari}}]{khan+2019}
{Khan}, S., {Miglio}, A., {Mosser}, B., {et~al.} 2019, \aap, 628, A35,
  \dodoi{10.1051/0004-6361/201935304}

\bibitem[{{Kjeldsen} \& {Bedding}(1995)}]{kjeldsen&bedding1995}
{Kjeldsen}, H., \& {Bedding}, T.~R. 1995, \aap, 293, 87

\bibitem[{{Kurucz}(1970)}]{kurucz+1970a}
{Kurucz}, R.~L. 1970, SAO Special Report, 309

\bibitem[{{Kurucz}(1993)}]{kurucz+1993a}
{Kurucz}, R.~L. 1993, in Astronomical Society of the Pacific Conference Series,
  Vol.~44, IAU Colloq. 138: Peculiar versus Normal Phenomena in A-type and
  Related Stars, ed. M.~M. {Dworetsky}, F.~{Castelli}, \& R.~{Faraggiana}, 87

\bibitem[{{Lindegren} {et~al.}(2018){Lindegren}, {Hern{\'a}ndez}, {Bombrun},
  {Klioner}, {Bastian}, {Ramos-Lerate}, {de Torres}, {Steidelm{\"u}ller},
  {Stephenson}, {Hobbs}, {Lammers}, {Biermann}, {Geyer}, {Hilger}, {Michalik},
  {Stampa}, {McMillan}, {Casta{\~n}eda}, {Clotet}, {Comoretto}, {Davidson},
  {Fabricius}, {Gracia}, {Hambly}, {Hutton}, {Mora}, {Portell}, {van Leeuwen},
  {Abbas}, {Abreu}, {Altmann}, {Andrei}, {Anglada}, {Balaguer-N{\'u}{\~n}ez},
  {Barache}, {Becciani}, {Bertone}, {Bianchi}, {Bouquillon}, {Bourda},
  {Br{\"u}semeister}, {Bucciarelli}, {Busonero}, {Buzzi}, {Cancelliere},
  {Carlucci}, {Charlot}, {Cheek}, {Crosta}, {Crowley}, {de Bruijne}, {de
  Felice}, {Drimmel}, {Esquej}, {Fienga}, {Fraile}, {Gai}, {Garralda},
  {Gonz{\'a}lez-Vidal}, {Guerra}, {Hauser}, {Hofmann}, {Holl}, {Jordan},
  {Lattanzi}, {Lenhardt}, {Liao}, {Licata}, {Lister}, {L{\"o}ffler},
  {Marchant}, {Martin-Fleitas}, {Messineo}, {Mignard}, {Morbidelli}, {Poggio},
  {Riva}, {Rowell}, {Salguero}, {Sarasso}, {Sciacca}, {Siddiqui}, {Smart},
  {Spagna}, {Steele}, {Taris}, {Torra}, {van Elteren}, {van Reeven}, \&
  {Vecchiato}}]{lindegren+2018}
{Lindegren}, L., {Hern{\'a}ndez}, J., {Bombrun}, A., {et~al.} 2018, \aap, 616,
  A2, \dodoi{10.1051/0004-6361/201832727}

\bibitem[{{Majewski} {et~al.}(2010){Majewski}, {Wilson}, {Hearty}, {Schiavon},
  \& {Skrutskie}}]{majewski+2010}
{Majewski}, S.~R., {Wilson}, J.~C., {Hearty}, F., {Schiavon}, R.~R., \&
  {Skrutskie}, M.~F. 2010, in IAU Symposium, Vol. 265, Chemical Abundances in
  the Universe: Connecting First Stars to Planets, ed. K.~{Cunha}, M.~{Spite},
  \& B.~{Barbuy}, 480--481, \dodoi{10.1017/S1743921310001298}

\bibitem[{{Mamajek} {et~al.}(2015){Mamajek}, {Prsa}, {Torres}, {Harmanec},
  {Asplund}, {Bennett}, {Capitaine}, {Christensen-Dalsgaard}, {Depagne},
  {Folkner}, {Haberreiter}, {Hekker}, {Hilton}, {Kostov}, {Kurtz}, {Laskar},
  {Mason}, {Milone}, {Montgomery}, {Richards}, {Schou}, \&
  {Stewart}}]{mamajek+2015a}
{Mamajek}, E.~E., {Prsa}, A., {Torres}, G., {et~al.} 2015, arXiv e-prints

\bibitem[{Mathur {et~al.}(2017)Mathur, Huber, Batalha, Ciardi, Bastien,
  Bieryla, Buchhave, Cochran, Endl, Esquerdo, Furlan, Howard, Howell, Isaacson,
  Latham, MacQueen, \& Silva}]{mathur+2017}
Mathur, S., Huber, D., Batalha, N.~M., {et~al.} 2017, The Astrophysical Journal
  Supplement Series, 229, 30

\bibitem[{McKinney(2010)}]{pandas}
McKinney, W. 2010, in Proceedings of the 9th Python in Science Conference, ed.
  S.~van~der Walt \& J.~Millman, 51 -- 56

\bibitem[{{Michalik} {et~al.}(2015){Michalik}, {Lindegren}, \&
  {Hobbs}}]{michalik_lindegren&hobbs2015}
{Michalik}, D., {Lindegren}, L., \& {Hobbs}, D. 2015, \aap, 574, A115,
  \dodoi{10.1051/0004-6361/201425310}

\bibitem[{{Mosser} {et~al.}(2013){Mosser}, {Dziembowski}, {Belkacem}, {Goupil},
  {Michel}, {Samadi}, {Soszy{\'n}ski}, {Vrard}, {Elsworth}, {Hekker}, \&
  {Mathur}}]{mosser+2013}
{Mosser}, B., {Dziembowski}, W.~A., {Belkacem}, K., {et~al.} 2013, \aap, 559,
  A137, \dodoi{10.1051/0004-6361/201322243}

\bibitem[{{Paxton} {et~al.}(2011){Paxton}, {Bildsten}, {Dotter}, {Herwig},
  {Lesaffre}, \& {Timmes}}]{paxton+2011a}
{Paxton}, B., {Bildsten}, L., {Dotter}, A., {et~al.} 2011, \apjs, 192, 3

\bibitem[{{Paxton} {et~al.}(2013){Paxton}, {Cantiello}, {Arras}, {Bildsten},
  {Brown}, {Dotter}, {Mankovich}, {Montgomery}, {Stello}, {Timmes}, \&
  {Townsend}}]{paxton+2013a}
{Paxton}, B., {Cantiello}, M., {Arras}, P., {et~al.} 2013, \apjs, 208, 4

\bibitem[{{Paxton} {et~al.}(2015){Paxton}, {Marchant}, {Schwab}, {Bauer},
  {Bildsten}, {Cantiello}, {Dessart}, {Farmer}, {Hu}, {Langer}, {Townsend},
  {Townsley}, \& {Timmes}}]{paxton+2015a}
{Paxton}, B., {Marchant}, P., {Schwab}, J., {et~al.} 2015, \apjs, 220, 15

\bibitem[{P{\'e}rez \& Granger(2007)}]{ipython}
P{\'e}rez, F., \& Granger, B.~E. 2007, Computing in Science \& Engineering, 9,
  21, \dodoi{10.1109/MCSE.2007.53}

\bibitem[{{Pinsonneault} {et~al.}(2012){Pinsonneault}, {An},
  {Molenda-{\.Z}akowicz}, {Chaplin}, {Metcalfe}, \&
  {Bruntt}}]{pinsonneault+2012a}
{Pinsonneault}, M.~H., {An}, D., {Molenda-{\.Z}akowicz}, J., {et~al.} 2012,
  \apjs, 199, 30, \dodoi{10.1088/0067-0049/199/2/30}

\bibitem[{{Pinsonneault} {et~al.}(2014){Pinsonneault}, {Elsworth}, {Epstein},
  {Hekker}, {M{\'e}sz{\'a}ros}, {Chaplin}, {Johnson}, {Garc{\'{\i}}a},
  {Holtzman}, {Mathur}, {Garc{\'{\i}}a P{\'e}rez}, {Silva Aguirre}, {Girardi},
  {Basu}, {Shetrone}, {Stello}, {Allende Prieto}, {An}, {Beck}, {Beers},
  {Bizyaev}, {Bloemen}, {Bovy}, {Cunha}, {De Ridder}, {Frinchaboy},
  {Garc{\'{\i}}a-Hern{\'a}ndez}, {Gilliland}, {Harding}, {Hearty}, {Huber},
  {Ivans}, {Kallinger}, {Majewski}, {Metcalfe}, {Miglio}, {Mosser}, {Muna},
  {Nidever}, {Schneider}, {Serenelli}, {Smith}, {Tayar}, {Zamora}, \&
  {Zasowski}}]{pinsonneault+2014}
{Pinsonneault}, M.~H., {Elsworth}, Y., {Epstein}, C., {et~al.} 2014, \apjs,
  215, 19, \dodoi{10.1088/0067-0049/215/2/19}

\bibitem[{{Pinsonneault} {et~al.}(2018){Pinsonneault}, {Elsworth}, {Tayar},
  {Serenelli}, {Stello}, {Zinn}, {Mathur}, {Garc{\'{\i}}a}, {Johnson},
  {Hekker}, {Huber}, {Kallinger}, {M{\'e}sz{\'a}ros}, {Mosser}, {Stassun},
  {Girardi}, {Rodrigues}, {Silva Aguirre}, {An}, {Basu}, {Chaplin}, {Corsaro},
  {Cunha}, {Garc{\'{\i}}a-Hern{\'a}ndez}, {Holtzman}, {J{\"o}nsson},
  {Shetrone}, {Smith}, {Sobeck}, {Stringfellow}, {Zamora}, {Beers},
  {Fern{\'a}ndez-Trincado}, {Frinchaboy}, {Hearty}, \&
  {Nitschelm}}]{pinsonneault+2018}
{Pinsonneault}, M.~H., {Elsworth}, Y.~P., {Tayar}, J., {et~al.} 2018, \apjs,
  239, 32, \dodoi{10.3847/1538-4365/aaebfd}

\bibitem[{{Rawls} {et~al.}(2016){Rawls}, {Gaulme}, {McKeever}, {Jackiewicz},
  {Orosz}, {Corsaro}, {Beck}, {Mosser}, {Latham}, \& {Latham}}]{rawls+2016a}
{Rawls}, M.~L., {Gaulme}, P., {McKeever}, J., {et~al.} 2016, \apj, 818, 108

\bibitem[{{Rodrigues} {et~al.}(2014){Rodrigues}, {Girardi}, {Miglio},
  {Bossini}, {Bovy}, {Epstein}, {Pinsonneault}, {Stello}, {Zasowski}, {Allende
  Prieto}, {Chaplin}, {Hekker}, {Johnson}, {M{\'e}sz{\'a}ros}, {Mosser},
  {Anders}, {Basu}, {Beers}, {Chiappini}, {da Costa}, {Elsworth},
  {Garc{\'{\i}}a}, {Garc{\'{\i}}a P{\'e}rez}, {Hearty}, {Maia}, {Majewski},
  {Mathur}, {Montalb{\'a}n}, {Nidever}, {Santiago}, {Schultheis}, {Serenelli},
  \& {Shetrone}}]{rodrigues+2014}
{Rodrigues}, T.~S., {Girardi}, L., {Miglio}, A., {et~al.} 2014, \mnras, 445,
  2758, \dodoi{10.1093/mnras/stu1907}

\bibitem[{{Sahlholdt} \& {Silva Aguirre}(2018)}]{sahlholdt+2018b}
{Sahlholdt}, C.~L., \& {Silva Aguirre}, V. 2018, \mnras, 481, L125

\bibitem[{{Salaris} {et~al.}(1993){Salaris}, {Chieffi}, \&
  {Straniero}}]{salaris+1993a}
{Salaris}, M., {Chieffi}, A., \& {Straniero}, O. 1993, \apj, 414, 580

\bibitem[{{Schlafly} \& {Finkbeiner}(2011)}]{sf2011}
{Schlafly}, E.~F., \& {Finkbeiner}, D.~P. 2011, \apj, 737, 103,
  \dodoi{10.1088/0004-637X/737/2/103}

\bibitem[{{Schlegel} {et~al.}(1998){Schlegel}, {Finkbeiner}, \&
  {Davis}}]{sfd1998}
{Schlegel}, D.~J., {Finkbeiner}, D.~P., \& {Davis}, M. 1998, \apj, 500, 525,
  \dodoi{10.1086/305772}

\bibitem[{{Serenelli} {et~al.}(2017){Serenelli}, {Johnson}, {Huber},
  {Pinsonneault}, {Ball}, {Tayar}, {Silva Aguirre}, {Basu}, {Troup}, {Hekker},
  {Kallinger}, {Stello}, {Davies}, {Lund}, {Mathur}, {Mosser}, {Stassun},
  {Chaplin}, {Elsworth}, {Garc{\'{\i}}a}, {Handberg}, {Holtzman}, {Hearty},
  {Garc{\'{\i}}a-Hern{\'a}ndez}, {Gaulme}, \& {Zamora}}]{serenelli17}
{Serenelli}, A., {Johnson}, J., {Huber}, D., {et~al.} 2017, \apjs, 233, 23,
  \dodoi{10.3847/1538-4365/aa97df}

\bibitem[{{Serenelli} {et~al.}(2013){Serenelli}, {Bergemann}, {Ruchti}, \&
  {Casagrande}}]{serenelli+2013a}
{Serenelli}, A.~M., {Bergemann}, M., {Ruchti}, G., \& {Casagrande}, L. 2013,
  \mnras, 429, 3645

\bibitem[{{Sharma} \& {Stello}(2016)}]{asfgrid}
{Sharma}, S., \& {Stello}, D. 2016, {Asfgrid: Asteroseismic parameters for a
  star}.
\newblock \doeprint{1603.009}

\bibitem[{{Sharma} {et~al.}(2016){Sharma}, {Stello}, {Bland-Hawthorn}, {Huber
  }, \& {Bedding}}]{sharma+2016}
{Sharma}, S., {Stello}, D., {Bland-Hawthorn}, J., {Huber }, D., \& {Bedding},
  T.~R. 2016, \apj, 822, 15, \dodoi{10.3847/0004-637X/822/1/15}

\bibitem[{{Silva Aguirre} {et~al.}(2012){Silva Aguirre}, {Casagrande}, {Basu},
  {Campante}, {Chaplin}, {Huber}, {Miglio}, {Serenelli}, {Ballot}, {Bedding},
  {Christensen-Dalsgaard}, {Creevey}, {Elsworth}, {Garc{\'{\i}}a}, {Gilliland},
  {Hekker}, {Kjeldsen}, {Mathur}, {Metcalfe}, {Monteiro}, {Mosser},
  {Pinsonneault}, {Stello}, {Weiss}, {Tenenbaum}, {Twicken}, \&
  {Uddin}}]{silva_aguirre+2012}
{Silva Aguirre}, V., {Casagrande}, L., {Basu}, S., {et~al.} 2012, \apj, 757,
  99, \dodoi{10.1088/0004-637X/757/1/99}

\bibitem[{{Skrutskie} {et~al.}(2006){Skrutskie}, {Cutri}, {Stiening},
  {Weinberg}, {Schneider}, {Carpenter}, {Beichman}, {Capps}, {Chester},
  {Elias}, {Huchra}, {Liebert}, {Lonsdale}, {Monet}, {Price}, {Seitzer},
  {Jarrett}, {Kirkpatrick}, {Gizis}, {Howard}, {Evans}, {Fowler}, {Fullmer},
  {Hurt}, {Light}, {Kopan}, {Marsh}, {McCallon}, {Tam}, {Van Dyk}, \&
  {Wheelock}}]{skrutskie+2006}
{Skrutskie}, M.~F., {Cutri}, R.~M., {Stiening}, R., {et~al.} 2006, \aj, 131,
  1163, \dodoi{10.1086/498708}

\bibitem[{{Stassun} {et~al.}(2017){Stassun}, {Collins}, \&
  {Gaudi}}]{stassun+2017a}
{Stassun}, K.~G., {Collins}, K.~A., \& {Gaudi}, B.~S. 2017, \aj, 153, 136

\bibitem[{{Stassun} \& {Torres}(2016)}]{stassun+2016a}
{Stassun}, K.~G., \& {Torres}, G. 2016, \aj, 152, 180

\bibitem[{{Stello} {et~al.}(2014){Stello}, {Compton}, {Bedding},
  {Christensen-Dalsgaard}, {Kiss}, {Kjeldsen}, {Bellamy}, {Garc{\'{\i}}a}, \&
  {Mathur}}]{stello+2014}
{Stello}, D., {Compton}, D.~L., {Bedding}, T.~R., {et~al.} 2014, \apjl, 788,
  L10, \dodoi{10.1088/2041-8205/788/1/L10}

\bibitem[{{Tassoul}(1980)}]{tassoul1980}
{Tassoul}, M. 1980, \apjs, 43, 469, \dodoi{10.1086/190678}

\bibitem[{{Torres}(2010)}]{torres+2010a}
{Torres}, G. 2010, \aj, 140, 1158

\bibitem[{{van Leeuwen}(2007)}]{vanleeuwen2007}
{van Leeuwen}, F. 2007, \aap, 474, 653, \dodoi{10.1051/0004-6361:20078357}

\bibitem[{{Viani} {et~al.}(2017){Viani}, {Basu}, {Chaplin}, {Davies}, \&
  {Elsworth}}]{viani+2017a}
{Viani}, L.~S., {Basu}, S., {Chaplin}, W.~J., {Davies}, G.~R., \& {Elsworth},
  Y. 2017, \apj, 843, 11

\bibitem[{Walt {et~al.}(2011)Walt, Colbert, \& Varoquaux}]{numpy}
Walt, S. v.~d., Colbert, S.~C., \& Varoquaux, G. 2011, Computing in Science \&
  Engineering, 13, 22, \dodoi{10.1109/MCSE.2011.37}

\bibitem[{White {et~al.}(2011)White, Bedding, Stello, Christensen-Dalsgaard,
  Huber, \& Kjeldsen}]{white+2011}
White, T.~R., Bedding, T.~R., Stello, D., {et~al.} 2011, The Astrophysical
  Journal, 743, 161

\bibitem[{{White} {et~al.}(2013){White}, {Huber}, {Maestro}, {Bedding},
  {Ireland}, {Baron}, {Boyajian}, {Che}, {Monnier}, {Pope}, {Roettenbacher},
  {Stello}, {Tuthill}, {Farrington}, {Goldfinger}, {McAlister}, {Schaefer},
  {Sturmann}, {Sturmann}, {ten Brummelaar}, \& {Turner}}]{white13}
{White}, T.~R., {Huber}, D., {Maestro}, V., {et~al.} 2013, \mnras, 433, 1262,
  \dodoi{10.1093/mnras/stt802}

\bibitem[{{Yu} {et~al.}(2018){Yu}, {Huber}, {Bedding}, {Stello}, {Hon},
  {Murphy}, \& {Khanna}}]{yu+2018}
{Yu}, J., {Huber}, D., {Bedding}, T.~R., {et~al.} 2018, \apjs, 236, 42,
  \dodoi{10.3847/1538-4365/aaaf74}

\bibitem[{{Zinn} {et~al.}(2017){Zinn}, {Huber}, {Pinsonneault}, \&
  {Stello}}]{zinn+2017a}
{Zinn}, J.~C., {Huber}, D., {Pinsonneault}, M.~H., \& {Stello}, D. 2017, \apj,
  844, 166

\bibitem[{{Zinn} {et~al.}(2019){Zinn}, {Pinsonneault}, {Huber}, \&
  {Stello}}]{zinn+2019}
{Zinn}, J.~C., {Pinsonneault}, M.~H., {Huber}, D., \& {Stello}, D. 2019, \apj,
  878, 136, \dodoi{10.3847/1538-4357/ab1f66}

\end{thebibliography}

\label{lastpage}
\end{document}